\documentclass[useAMS,usenatbib,aas_macros,fleqn]{mn2e}

\usepackage{verbatim}

\usepackage{epsfig}
\usepackage{amssymb}
\usepackage{natbib}
\usepackage{times}

\usepackage{multirow}
\usepackage{xcolor}

\usepackage{graphics}
\usepackage{amsfonts}
\usepackage{amsmath}
\usepackage{layout}
\usepackage{amssymb}
\usepackage{epsfig}

\usepackage{setspace}

\usepackage{hyperref}

\usepackage{etoolbox}
\makeatletter
\newcount\c@additionalboxlevel
\newcount\c@maxboxlevel
\patchcmd\@combinedblfloats{\box\@outputbox}{%
  \stepcounter{additionalboxlevel}%
  \box\@outputbox
}{}{\errmessage{\noexpand\@combinedblfloats could not be patched}}

\AtBeginShipout{%
  \ifnum\value{additionalboxlevel}>\value{maxboxlevel}%
    \typeout{Warning: maxboxlevel might be too small, increase to %
      \the\value{additionalboxlevel}%
    }%
  \fi
  \@whilenum\value{additionalboxlevel}<\value{maxboxlevel}\do{%
    \typeout{* Additional boxing of page `\thepage'}%
    \setbox\AtBeginShipoutBox=\hbox{\copy\AtBeginShipoutBox}%
    \stepcounter{additionalboxlevel}%
  }%
  \setcounter{additionalboxlevel}{0}%
}
\makeatother 

\newcommand{\figu}{Figure~}

\newcommand{\eq}{Equation~}

\newcommand{\sect}{Section~}

\newcommand{\ergse}{{\rm erg\, s^{-1}}}

\newcommand{\sis}{$\sigma$}
\newcommand{\sise}{\sigma}

\newcommand{\mbh}{$M_{\rm bh}$}
\newcommand{\mbhe}{M_{\rm bh}}

\newcommand{\mstar}{$M_{\rm star}$}
\newcommand{\mstare}{M_{\rm star}}

\newcommand{\msune}{M_{\odot}}

\newcommand{\kbe}{k_{\rm bol}}
\newcommand{\kb}{$k_{\rm bol}$}

\newcommand{\epsie}{\varepsilon}
\newcommand{\epsi}{$\varepsilon$}

\newcommand{\macce}{\dot{M}_{\rm BH,acc}}

\newcommand{\fvire}{f_{vir}}

\begin{document}

\voffset=-0.50in

\def\sarc{$^{\prime\prime}\!\!.$}
\def\arcsec{$^{\prime\prime}$}
\def\arcmin{$^{\prime}$}
\def\degr{$^{\circ}$}
\def\seco{$^{\rm s}\!\!.$}
\def\ls{\lower 2pt \hbox{$\;\scriptscriptstyle \buildrel<\over\sim\;$}}
\def\gs{\lower 2pt \hbox{$\;\scriptscriptstyle \buildrel>\over\sim\;$}}

\title[Black growth from stacked X-ray AGN]{Probing black hole accretion tracks, scaling relations and radiative efficiencies from stacked X-ray active galactic nuclei}

\author[F. Shankar et al.]
{Francesco Shankar$^{1}$\thanks{E-mail:$\;$F.Shankar@soton.ac.uk}, David H. Weinberg$^{2}$, Christopher Marsden$^{1}$, Philip J. Grylls$^{1}$, \newauthor
Mariangela Bernardi$^{3}$, Guang Yang$^{4}$, Benjamin Moster$^{5}$, Rosamaria Carraro$^{6}$, David M. \newauthor
Alexander$^{7}$, Viola Allevato$^{8}$, Tonima T. Ananna$^{9}$, Angela Bongiorno$^{10}$, Giorgio Calderone$^{11}$, \newauthor
Francesca Civano$^{12}$,  Emanuele Daddi$^{13}$, Ivan DelVecchio$^{13}$, Federica Duras$^{14}$, Fabio \newauthor
La Franca$^{14}$, Andrea Lapi$^{15}$, Youjun Lu$^{16}$, Nicola Menci$^{10}$, Mar Mezcua$^{17}$, Federica Ricci$^{18}$, \newauthor
Giulia Rodighiero$^{19}$, Ravi K. Sheth$^{3}$, Hyewon Suh$^{20}$, Carolin Villforth$^{21}$, Lorenzo Zanisi$^{1}$
\\
$1$ Department of Physics and Astronomy, University of Southampton, Highfield, SO17 1BJ, UK\\
$2$ Department of Astronomy and the Center for Cosmology and Astroparticle Physics, The Ohio State University,
       140 West 18th Avenue, Columbus, OH 43210, USA\\
$3$ Department of Physics and Astronomy, University of Pennsylvania, 209 South 33rd St, Philadelphia, PA 19104\\
$4$ Department of Physics and Astronomy, Texas A\&M University, College Station, TX 77843-4242, USA\\
$5$ Universit\"{a}ts-Sternwarte, Ludwig-Maximilians-Universit\"{a}t M\"{u}nchen, Scheinerstr. 1, 81679 M\"{u}nchen, Germany\\
$6$ Instituto de F\'\i{}sica y Astronom\'\i{}a, Universidad de Valpara\'\i{}so, Gran Bretaña 1111, Playa Ancha, Valpara\'\i{}so, Chile\\
$7$ Centre for Extragalactic Astronomy, Department of Physics, Durham University, South Road, Durham DH1 3LE, UK\\
$8$ Scuola Normale Superiore, Piazza dei Cavalieri 7, I-56126 Pisa, Italy\\
$9$ Department of Physics, Yale University, P.O. Box 201820, New Haven, CT 06520-8120, USA\\
$10$ INAFOsservatorio Astronomico di Roma, via Frascati 33, 00078 Monteporzio Catone, Italy\\
$11$ INAF  Osservatorio Astronomico di Trieste, Via Tiepolo 11, I-34131 Trieste, Italy\\
$12$ Harvard-Smithsonian Center for Astrophysics, Cambridge, MA, 02138, USA\\
$13$ CEA, IRFU, DAp, AIM, Universit\'{e} Paris-Saclay, Universit\'{e} Paris Diderot, Sorbonne Paris Cit\'{e}, CNRS, F-91191 Gif-sur-Yvette,
France\\
$14$ Dipartimento di Matematica e Fisica, Universit\`{a} Roma Tre, via della Vasca Navale 84, I-00146 Roma, Italy\\
$15$ SISSA, Via Bonomea 265, I-34136 Trieste, Italy\\
$16$ National Astronomical Observatories, Chinese Academy of Sciences, Beijing 100101, Peoples Republic of China\\
$17$ Institute of Space Sciences (ICE, CSIC), Campus UAB, Carrer de Magrans, E-08193 Barcelona, Spain\\
$18$ Instituto de Astrof\'{\i}sica and Centro de Astroingenier\'{\i}a, Facultad de F\'{\i}sica, Pontificia Universidad Cat\'{o}lica de Chile, Casilla 306, Santiago 22, Chile\\
$19$ University of Padova, Physics and Astronomy Department, Vicolo Osservatorio 3, 35122, Padova, Italy\\
$20$ Subaru Telescope, National Astronomical Observatory of Japan (NAOJ), 650 North Aohoku place, Hilo, HI 96720,
USA\\
$21$ Department of Physics, University of Bath, Claverton Down, BA2 7AY Bath, UK\\
}
\date{}
\pagerange{\pageref{firstpage}--
\pageref{lastpage}} \pubyear{2019}
\maketitle
\label{firstpage}

\begin{abstract}
The masses of supermassive black holes at the centres of local galaxies appear to be tightly correlated with the mass and velocity dispersions of their galactic hosts. However, the local \mbh-\mstar\ relation inferred from dynamically measured inactive black holes is up to an order-of-magnitude higher than some estimates from active black holes, and recent work suggests that this discrepancy arises from selection bias on the sample of dynamical black hole mass measurements. In this work we combine X-ray measurements of the mean black hole accretion luminosity as a function of stellar mass and redshift with empirical models of galaxy stellar mass growth, integrating over time to predict the evolving \mbh-\mstar\ relation. The implied relation is nearly independent of redshift, indicating that stellar and black hole masses grow, on average, at similar rates. Matching the de-biased local \mbh-\mstar\ relation requires a mean radiative efficiency $\epsie \gtrsim 0.15$, in line with theoretical expectations for accretion onto spinning black holes. However, matching the ``raw'' observed relation for inactive black holes requires $\epsie \sim 0.02$, far below theoretical expectations. This result provides independent evidence for selection bias in dynamically estimated black hole masses, a conclusion that is robust to uncertainties in bolometric corrections, obscured active black hole fractions, and kinetic accretion efficiency. For our fiducial assumptions, they favour moderate-to-rapid spins of typical supermassive black holes, to achieve $\epsie\sim 0.12-0.20$. Our approach has similarities to the classic Soltan analysis, but by using galaxy-based data instead of integrated quantities we are able to focus on regimes where observational uncertainties are minimized.
\end{abstract}


\section{Introduction}
\label{sec|intro}

Supermassive black holes are detected at the centres of almost all local galaxies observed with high enough sensitivity, and they seem to share close links with their host galaxies. The mass of central black holes is observed to scale proportionally with the stellar mass of the host galaxy and with the fourth or fifth power of its stellar velocity dispersion \citep[e.g.,][]{Magorrian98,Ferrarese00,Marconi04,HaringRix,KormendyHo,Laesker14,GrahamScott15,Remco15,Savorgnan16,Shankar16BH,Sahu19},
suggesting a ``co-evolution'' between the black holes and their hosts \citep[e.g.,][]{Granato04,Lapi06,Shankar06,Hop08clust}. In particular, from analysis of the residuals in the various scaling relations, evidence was put forward that black hole mass \mbh\ is mostly correlated to velocity dispersion \sis, rather than stellar mass \mstar\ or any other galactic property \citep[e.g.,][]{Bernardi07,Shankar17BH,Shankar19BH,deNicolaMarconi19}, a possible signature of momentum/energetic feedback from the central black hole on their hosts during their bright phases as active galactic nuclei (AGN) \citep[e.g.,][]{SilkRees,King03,Fabian12,Zubovas19}. In this context, a correlation between black hole mass and host galaxy (total) stellar mass would then be a byproduct of the more fundamental \mbh-\sis\ and \sis-\mstar\ relations.

\begin{figure*}
    \center{\includegraphics[width=\textwidth]{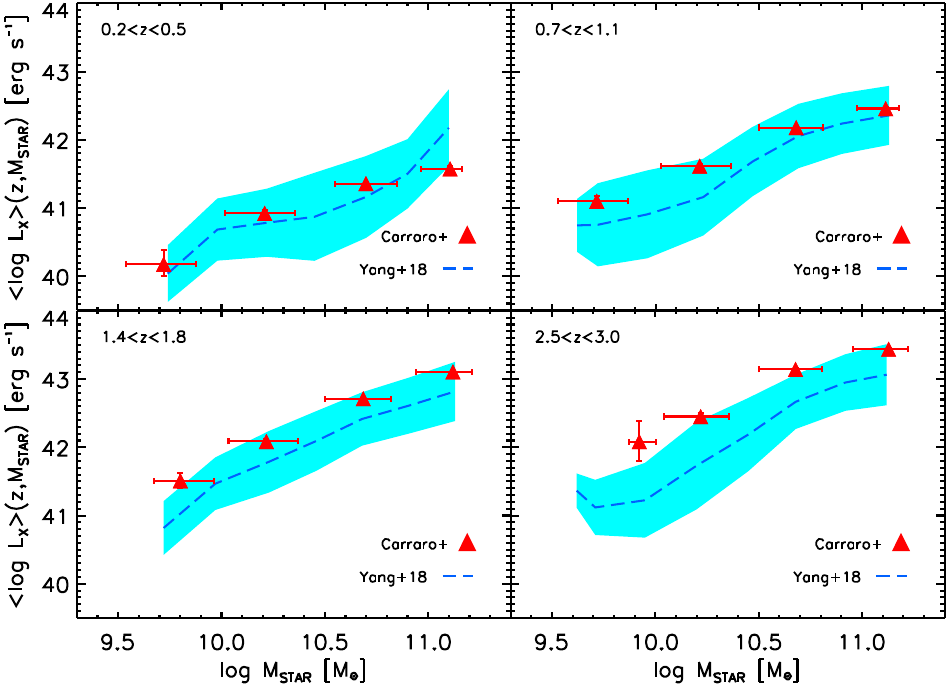}
    \caption{X-ray luminosities, averaged over active and inactive galaxies, as a function of stellar mass and redshift. Data are from \citet{Yang18} and Carraro et al. (2019, submitted), as labelled. The cyan region brackets the 1$\sigma$ scatter around the mean.
    \label{fig|FigureBHacc}}}
\end{figure*}

Deciphering the origin and evolution of supermassive black holes in galaxies requires proper observational characterization of the black hole-galaxy scaling relations, which however remains a non-trivial challenge. One of the most pressing issues in this respect is the possible presence of observational biases affecting the scaling relations \citep[e.g.,][]{YuTremaine,Bernardi07,Batcheldor07,Gultekin11,DaiMbhSigma,Shankar16BH}. Following the preliminary work by \citet{Bernardi07}, \citet{Shankar16BH} more recently emphasized that samples of local quiescent (mainly early-type) galaxies having dynamically-measured central black hole masses present larger velocity dispersions at fixed stellar mass with respect to the mean trend for early-type galaxies in the Sloan Digital Sky Survey (SDSS). Via targeted Monte Carlo simulations in which black hole mass was assumed to scale as $\mbhe \propto \sise^{4-5}$, \citet{Shankar16BH} showed that the apparent discrepancies in the velocity distributions at fixed stellar mass could be straightforwardly explained in terms of an observational selection effect. To perform reliable dynamical black hole mass measurements, the black hole gravitational sphere of influence\footnote{In the $r_g$ formula in the text, the velocity dispersion is calculated at large scales, outside of the gravitational sphere of influence of the central black hole, and the constant of proportionality takes into account the galaxy profile. Discussions can be found in \citet{Shankar16BH} and \citet{Barausse17}.}, $r_g\propto \mbhe/\sise^2\propto \sise^{\beta}$ with $\beta\sim 2-3$, must be sufficiently resolved \citep[e.g.,][]{FerrareseFord}. The limited capabilities of present-day telescopes will inevitably favour the galaxies with the largest gravitational radii $r_g$, thus the highest velocity dispersions and highest black hole mass at fixed host galaxy stellar mass, biasing the observed scaling relations towards fictitiously higher normalizations. The Monte Carlo simulations showed that this gravitational bias by itself could account for the whole observed discrepancies in velocity dispersion distributions between SDSS galaxies and galaxies with dynamically-measured black holes, while predicting biases up to an order of magnitude in the observed \mbh-\mstar\ relation. In what follows, we will always refer to the directly observed \mbh-\mstar\ relation as ``raw'', and the claimed intrinsic \mbh-\mstar\ relation from \citet{Shankar16BH} as ``de-biased''. We will draw on additional observations and theoretical expectations of black hole accretion efficiency to argue that the de-biased relations are indeed more accurate.
In a recent conference proceedings, \citet{Kormendy19} has argued that the scaling relations derived from dynamically measured black holes \citep[e.g.,][]{KormendyHo}
are not biased; we address each of the points raised in his article in Appendix\ref{Appendix}.

AGN samples with reverberation or single-epoch black hole mass estimates do not suffer from the restriction of needing to observationally resolve the (small) central black hole gravitational sphere of influence, as their black hole masses are retrieved from the virial product of the broad emission line dispersions, which trace the gravitational potential in a region dominated by the black hole, and the radii inferred directly from reverberation mappings or indirectly from the size-luminosity relation \citep[e.g.,][]{Peterson04,Bentz08RL}. If local AGN are random samples of the underlying population of dynamically-measured supermassive black holes, they would be naturally expected to more closely trace the intrinsic/de-biased, rather than the observed/raw, \mbh-\mstar\ relation \citep{Shankar19BH}. Indeed, several groups found clear evidence for AGN to lie below the \mbh-\mstar\ relation of local, inactive black holes \citep[e.g.,][]{busch2014low,Dasyra07,Kim08,Sarria10,Falomo14,ReinesVolonteri15,greene2016megamaser,Ricci17SR,Bentz18,Shankar19BH}, when adopting virial factors $\fvire \sim 4$ as suggested by geometric and dynamic modelling of the broad line region \citep[e.g.,][]{Pancoast14,Grier17}.
More recently Shankar et al. (2019b, submitted) showed that the large-scale clustering as a function of black hole mass, as measured at $z=0.25$ from large-scale optical and X-ray surveys by \citet{Krumpe15}, is fully consistent with the de-biased, rather than the raw, local \mbh-\mstar\ relation, further suggesting the presence of a bias in the latter.

The central aim of the present work is to probe the shape, normalization and evolution of the relation between black hole mass and host galaxy (total) stellar mass \mbh-\mstar\ relation, in ways \emph{independent} of the local sample of dynamically-measured supermassive black holes. To this purpose, following the seminal works by \citet[e.g.,][]{Mulla12} and, in particular, \citet{Yang18}, we compute the \mbh-\mstar\ relation and its evolution with redshift adopting a new methodology which relies on large and deep X-ray AGN samples. More specifically, adopting the standard assumption that supermassive black holes are the relics of single or multiple gas accretion episodes in AGN \citep{Lynden69,Rees84,Soltan} and that their luminous outputs are regulated by a radiative efficiency \epsi\ \citep[e.g.,][]{Bardeen72,Thorne74}, we can directly convert the average AGN luminosities of a population of galaxies into the average mass accretion rates of their black holes
$\langle \macce \rangle[\mstare(z),z] \propto \langle L_X[z,\mstare(z)] \rangle/ \epsie $. The average here includes those galaxies whose central black holes are inactive at a given time and thus contribute negligibly to the mean AGN luminosity.

By following the host stellar mass evolutionary tracks $\mstare(z)$, derived from state-of-the-art
semi-empirical models \citep[e.g.,][]{Moster18,Behroozi18,Grylls19}, we can integrate in time the mean accretion rate $\langle \macce \rangle[\mstare(z),z]$ to infer the mean black hole mass
$\langle\mbhe(z)\rangle$ at the centre of the host galaxy with average stellar mass $\langle\mstare(z)\rangle$, and thus build the mean $\langle\mbhe(z)\rangle-\langle\mstare(z)\rangle$ relation at all accessible cosmic epochs (mostly $z\lesssim 3$).
In our approach it is irrelevant whether stellar mass is a primary or secondary galaxy property related to  black hole mass, as it is simply adopted as a ``tracer'' of the central AGN activity through cosmic time.
Galaxy and black hole mergers are a potential complication to this approach but we will show that they should have little impact for the intermediate mass galaxies from which we derive our main constraints.

We will show that the method outlined above produces \mbh-\mstar\ relations at the present epoch in close agreement with the de-biased \mbh-\mstar\ relation when adopting reasonable values of $\epsie \gtrsim 0.1$, as expected from standard accretion disk theory \citep{SSdisk} and as inferred from direct UV spectral energy distribution (SED) fitting \citep[e.g.,][]{DavisLaor11,Capellupo15}.
On the other hand, matching the raw \mbh-\mstar\ relation would require unrealistically low radiative efficiencies of $\epsie \lesssim 0.04$. On the assumption of a time-invariant mean radiative efficiency, the results put forward in this work also point to a constant \mbh-\mstar\ relation at all cosmic epochs probed by the stacked X-ray data, in line with recent independent estimates of the \mbh-\mstar\ relation from high-redshift single-epoch AGN samples (Suh et al. 2019, submitted).

The method outlined in his work is similar in principle to the classical Soltan-type approach \citep{Soltan}, in which the mean radiative efficiency $\epsie$ is constrained by comparing the time-integrated accreted mass from (all) AGN, which scales with the (inverse) mean radiative efficiency, with the local supermassive black hole mass density or mass function \citep{Salucci99,YuTremaine,Marconi04,Shankar04,YuLu08,SWM,Shankar13,Aversa15,ZhangLu19eta}. A disadvantage of this classical approach is that it relies on integrated quantities, so it is sensitive to uncertainties at the extremes of the AGN luminosity function or black hole mass function \citep[see, e.g.,][for reviews]{ShankarReview,GrahamReview15}. The inference of the local black hole mass density is also sensitive to the uncertain scatter about the mean black hole-galaxy scaling relations. While some systematic uncertainties also affect the approach used here, we are able to focus on specific regimes of galaxy mass and AGN luminosity where these uncertainties are minimized.

The paper is organised as follows. In \sect\ref{sec|data} we briefly present the data we adopt as input to our calculations. Our methodology is then detailed in \sect\ref{sec|Method}. We provide our results in \sect\ref{sec|results} and conclude in \sect\ref{sec|discu}.
In what follows, wherever relevant we will adopt a reference cosmology with $h=0.7$, $\Omega_m=0.3$, $\Omega_{\Lambda}=0.7$, and a \citet{Chabrier03} stellar initial mass function (IMF).

\section{Data}
\label{sec|data}

\begin{figure}
    \center{\includegraphics[width=\columnwidth]{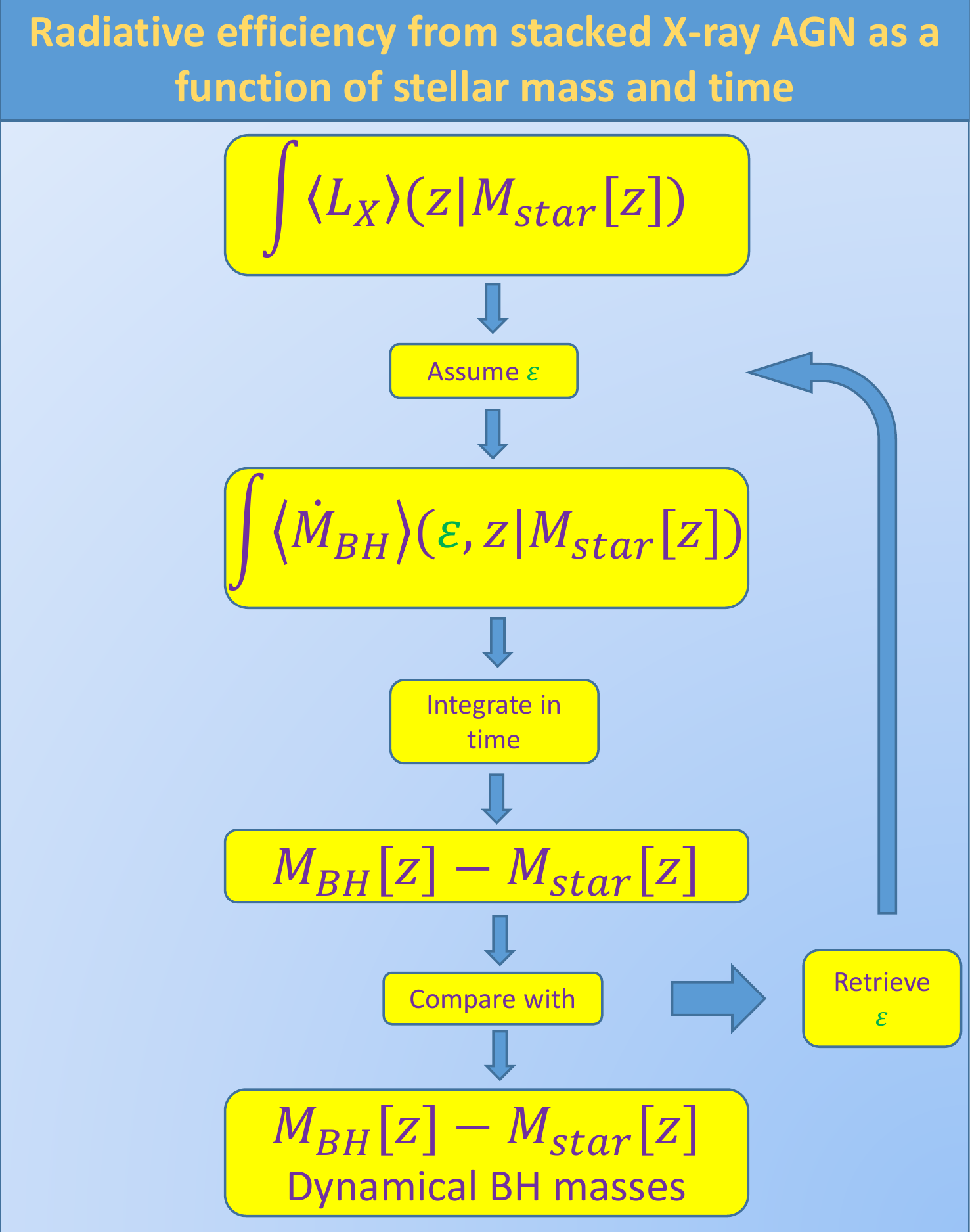}
    \caption{Cartoon visualizing the strategy of this work. After assuming a constant radiative efficiency \epsi, average black hole accretion histories are extracted from the X-ray luminosities as a function of host galaxy stellar mass and redshift averaged over the entire active and non active populations. We then follow input stellar mass growth histories $\mstare[z]$, which are converted to black hole mass accretion histories via $L_x(z,\mstare)$ and \epsi. The comparison with the local dynamically-based \mbh-\mstar\ relations (or at any redshift $z<2$ in which they are measured) can effectively constrain the input radiative efficiency, in ways largely independent of the obscured fraction of AGN (see text for details).
    \label{fig|SummaryCartoon}}}
\end{figure}

As our reference sample in this work we will make use of the X-ray luminosities from \citet{Yang18}, reported in \figu\ref{fig|FigureBHacc}. In brief, this sample of active and starforming galaxies has been extracted from the GOODS North and South and COSMOS galaxy samples with stellar masses derived from spectral energy distribution fitting of broad-band photometry\citep{Santini15}, cross-correlated with the Chandra Deep Fields North and South \citep[see][and references therein for full details]{Yang17}, assuming a \citet{Chabrier03} IMF and mass-to-light ratios computed as median among different methods, including \citet{BruzualCharlot03} and \citet{Maraston05}. Stellar masses in broad line AGN were further corrected by \citet{Yang18} to remove the AGN component.
Adding contributions from AGN in passive galaxies at each stellar mass would change results only slightly \citep{Yang18}.

Whilst \citep{Yang18} reference IMF is the same as the one adopted in this paper, their mass-to-light ratios, especially those by \citet{BruzualCharlot03}, may tend to provide less stellar mass than our reference \citet{Bell03SEDs} value, at fixed galaxy luminosity or colour \citep{Bell03SEDs}. Moreover, SED-based stellar masses may differ from photometrically-based ones, such as those adopted by \citet{Savorgnan16} and \citet{Shankar16BH} in deriving the host galaxy stellar masses of dynamically-measured local black holes. To check for systematic differences in stellar mass estimates, we have cross-correlated the low-redshift galaxies in \citet{Laigle16}, who make use of the SED fitting technique and \citet{BruzualCharlot03} mass-to-light ratios on the COSMOS field, with the photometrically-based stellar masses from the \citet{Meert15} catalogue, which was adopted as a reference by \citet{Shankar16BH}. We found the former to be, as expected, systematically smaller than the latter by a median of $\sim 0.15$ dex. To be conservative, we do not apply such a correction in our final estimates, noticing that increasing the final stellar masses of \citet{Yang18} at fixed black hole mass would if anything strengthen our main conclusions that reproducing the raw \mbh-\mstar\ relation requires a very low radiative efficiency.

To check on the accuracy of the luminosities computed by \citet{Yang18}, we compare their average X-ray luminosities as a function of stellar mass in \figu\ref{fig|FigureBHacc} (long-dashed, blue lines with cyan regions delimiting the 1$\sigma$ uncertainties) with data from Carraro et al. (submitted, red triangles), which have been extracted from Chandra stacking at 2-7 keV and converted to full band assuming $\Gamma=1.8$. We find very good agreement between the independent samples, supporting the validity of the \citet{Yang18} results. Averages in X-ray luminosity at a given stellar mass in \citet{Yang18} are taken over the full population of galaxies, including galaxies with no AGN detection. They are computed by full integration of the double power-law probability distributions $P(L_X|\mstare,z)$, which has been constrained from maximum-likelihood fitting by \citet{Yang18}. Such distributions have been shown, once convolved with the stellar mass function by \citet{Davidzon17}, to well reproduce the full X-ray luminosity function by \citet{Ueda14} at any redshift of interest.

\begin{figure*}
    \center{\includegraphics[width=\textwidth]{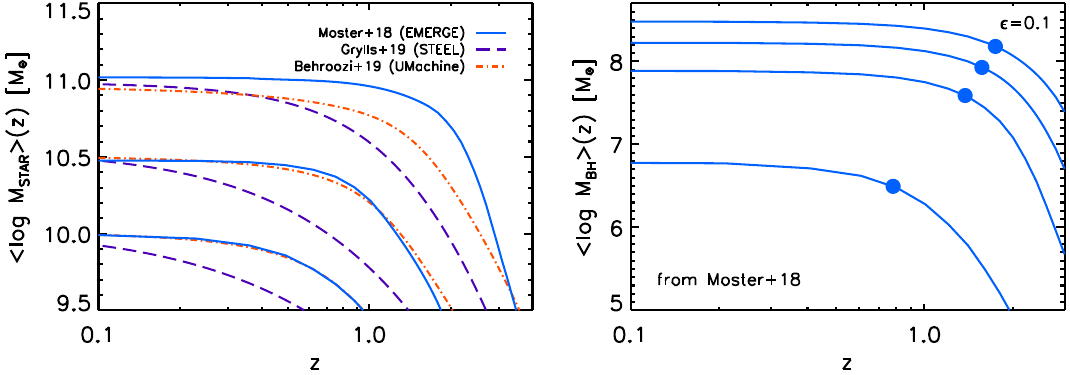}
    \caption{\emph{Left}: Examples of average stellar mass growth histories $\mstare[z]$ from \citet{Moster18}, \citet{Grylls19}, and \citet{Behroozi18}, as labelled. \emph{Right}: Examples of average black hole mass accretion histories $\mbhe[z]$ as expected from the mean X-ray luminosities of \figu\ref{fig|FigureBHacc}, assuming a radiative efficiency of $\epsie=0.1$ and the \citet{Moster18} stellar mass growth histories. The filled circle on each black hole mass track marks the redshift at which the black hole reaches 50\% of its final mass. Lower mass black holes gain more of their mass at late times, the behaviour often referred to as ``downsizing''.
    \label{fig|FigureMstarz}}}
\end{figure*}

\section{Method}
\label{sec|Method}

We here outline the step-by-step methodology pursued in this work to build black hole mass accretion histories and constrain mean radiative efficiencies. Our aim is to provide a novel framework that broadly builds upon the classical \citet{Soltan} argument, but also substantially expands beyond it making use of additional data and techniques.
As visualised in \figu\ref{fig|SummaryCartoon}, our approach consists of the following steps:
\begin{enumerate}
  \item We start from X-ray AGN luminosities converted to bolometric luminosities and averaged over the full populations of active and normal galaxies, and expressed as a function of stellar mass and redshift, $\langle L \rangle(\mstare,z)$.
  \item By assuming a mean radiative efficiency \epsi\ and kinetic efficiency $\epsie_{\rm kin}$, we convert average AGN bolometric luminosities into mean black hole mass accretion rates
      \begin{equation}
        \langle \macce \rangle(\mstare,z)=\frac{\langle L \rangle(\mstare,z) (1-\epsie-\epsie_{\rm kin})}{\epsie c^2} \, .
        \label{eq|MaccBH}
      \end{equation}
      The factor $(1-\epsie-\epsie_{\rm kin})$ in \eq\ref{eq|MaccBH} appears because \epsi\ is defined relative to the large scale accretion rate, but energy emitted as radiation or kinetic feedback does not contribute to the black hole's mass growth.
  \item We then make use of the mean galaxy mass accretion histories $\mstare[z]$, inferred from extensive cosmological semi-empirical models built around the abundance matching technique, to predict the average growth rates of supermassive black holes $\langle \macce \rangle(\mstare[z],z)$. Average mass growth histories of supermassive black holes are then simply built by integrating $\langle \macce \rangle(\mstare[z],z)$ along cosmic time.
  \item By integrating in redshift the galaxy and black hole mass accretion histories, we can retrieve the average black hole mass-stellar mass relation $\langle\mbhe[z]\rangle-\langle\mstare[z]\rangle$ at any redshift $z$ of interest\footnote{From now on, despite still referring to mean quantities throughout, we will usually drop the average symbols in black hole/galaxy stellar mass/accretion rates, for reasons of clarity.}. The comparison with the latest determination of the local \mbh-\mstar\ relation of dynamically-measured supermassive black holes will then constrain the mean radiative efficiency.
\end{enumerate}

The method outlined above is different from the traditional \citet{Soltan} approach as it does not deal with number densities but on mean accretion rates. It thus represents a novel, independent test of the connection between local black holes and distant AGN, and, as discussed below, it provides more robust constraints on the mean radiative efficiency of black holes. There are some key points important to emphasize at this stage. When comparing to a given rendition of the local \mbh-\mstar\ relation, we are actually constraining the \emph{ratio} between bolometric correction and radiative efficiency $\kbe(1-\epsie-\epsie_{\rm kin})/\epsie$. Nevertheless, we will see that within our current estimates of AGN bolometric corrections and obscured fractions, our proposed method provides a powerful test to bracket the allowed ranges of radiative efficiencies. We will also discuss the impact of allowing for additional kinetic losses in the estimate of the mean radiative efficiency.

\section{Results}
\label{sec|results}

\subsection{Average black hole mass accretion histories}
\label{subsec|aveMdotBH}

The first step of our modelling relies on computing reliable (average) X-ray luminosities as a function of stellar mass. As demonstrated by \citet{Yang17}, at fixed stellar mass any secondary dependence of X-ray luminosities on star formation rates are weak. It is thus a good approximation in what follows to consider, at any redshift of interest, only an explicit dependence of X-ray luminosities on total stellar mass. We note that more recently \citet[][see also \citealt{Ni19}]{Yang19} found evidence for a strong connection between X-ray luminosity and star formation rate when considering only the bulge component. However, our methodology does not necessarily rely on any causal connection between star formation and AGN activity or X-ray luminosity on galaxy stellar mass. Stellar mass growth tracks are simply used as ``tracers'' of AGN activity in our methodology, to connect descendants to progenitor AGN and thus estimate mean black hole accretion tracks.

X-ray luminosities averaged in small grids of redshift and stellar mass, are then converted to average black hole accretion rates as follows
\begin{equation}
\langle \macce \rangle=\int_{-2}^{\infty} P(L_X|\mstare,z) \\
\frac{(1-\epsie-\epsie_{\rm kin})\kbe L_X}{\epsie c^2}d\log L_X\, ,
    \label{eq|averageMacc}
\end{equation}
where \kb\ is the bolometric correction adapted from \citet[][see \figu8 in \citealt{Yang18}]{Lusso12}. The lower limit of integration $-2$ corresponds to a minimum \emph{specific} X-ray luminosity expressed in units of the host stellar mass. For a typical galaxy with mass $\mstare=10^{10}\, \msune$ this corresponds to an X-ray luminosity of $L_X\sim 4\times 10^{41}\ergse$ in the \citet{Yang18} AGN samples, sufficient to probe down to the faint end of the X-ray AGN luminosity function (see \citealt{Yang18} for full details). \citet{Yang18} performed additional tests to show that the cumulative black hole mass accreted at even lower specific X-ray luminosities is subdominant to the mass obtained via integration of \eq\ref{eq|averageMacc}. We also note that \eq\ref{eq|averageMacc} strictly holds at $0.4<z<4$, though, as already noted by \citet{Yang18}, extending the validity of \eq\ref{eq|averageMacc} to lower redshifts, as we do in the present work, adds a minor contribution to the final black hole mass.

\begin{figure*}
    \center{\includegraphics[width=\textwidth]{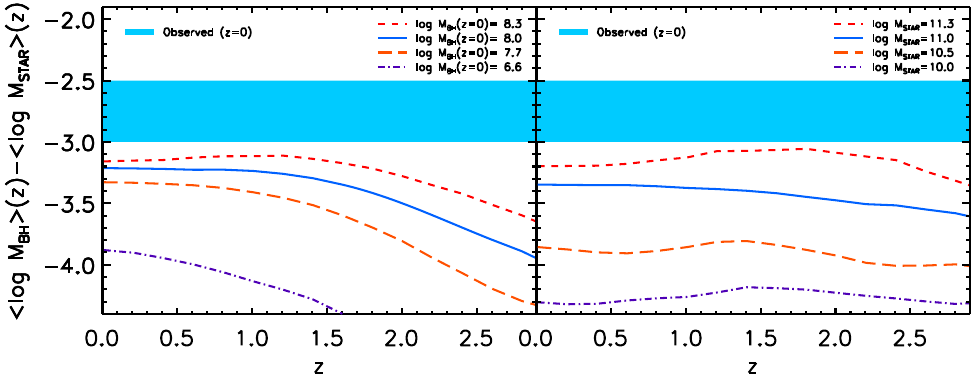}
    \caption{Examples of average black hole-to-stellar mass ratios as a function of redshift along the progenitors (left panel) and at fixed stellar mass (right panel), compared to the average ratio inferred by \citet{KormendyHo} in the local Universe (cyan region). At each \mstar\ in the right panel, the ratio $\langle\mbhe\rangle/\langle\mstare\rangle$ is roughly constant, at least for $z<2$.
    \label{fig|FigureMBHz}}}
\end{figure*}

To compute black hole mass accretion histories, we thus need reliable estimates of how the host galaxies actually grow in stellar mass. We here neglect any source of ``ex-situ'' accretion of stellar/black hole mass (e.g., mergers). This is a very good approximation as a number of cosmological analytic, semi-analytic, and numerical models \citep[e.g.,][]{Shankar13,Rodriguez17,Lapi18} agree in suggesting that the amount of stellar mass ex-situ is limited to $\lesssim 20\%$ for galaxies with $\mstare \lesssim (1-2)\times 10^{11}\, \msune$. \citet{Moster18}, \citet{Lapi18} and \citet{Grylls19b} have recently confirmed that, at least for the stellar mass range of interest to this work with $\log \mstare/\msune \lesssim 11.2$, the cumulative accretion via satellite mergers is limited to a few percent \citep[see also][]{Moster19}.

The left panel of \figu\ref{fig|FigureMstarz} shows the \citet{Moster18} semi empirically-constrained (\texttt{EMERGE} model) mean stellar mass growth histories of galaxies that today have a stellar mass of $\log\mstare[z=0]/\msune\sim 10, 10.5, 11$ (solid, blue lines), compared with another two recent semi-empirical models, the statistical model \texttt{STEEL} by \citet[][long-dashed purple lines]{Grylls19}, and the latest renditions of the \texttt{UniverseMachine} by
\citet[][dot-dashed, orange lines]{Behroozi18}. All of these models are based on tracking backwards or forward in time the host dark matter merger main progenitors, and at each time step computing the mass gained in mergers and lost due to stellar evolution given an input stellar mass-halo mass relation tuned to specifically reproduce the local stellar mass function of \citet{Bernardi13}. This function is based on the same stellar mass system adopted by \citet{Shankar16BH} and \citet{Shankar19BH} to retrieve the \mbh-\mstar\ relations adopted as a reference in this work. It is evident that despite being tuned against the same local stellar mass function, semi-empirical models may still produce noticeably distinct stellar mass growth tracks, with differences of up to $0.5$ dex at any given epoch. The origin of these discrepancies can, at least in part, be reconciled to differences in the high-redshift input observational data adopted by each group. For example, \citet{Moster18} tuned their model on larger star formation rates and lower stellar mass densities than those adopted in the \texttt{STEEL} reference model.

\begin{figure*}
    \center{\includegraphics[width=\textwidth]{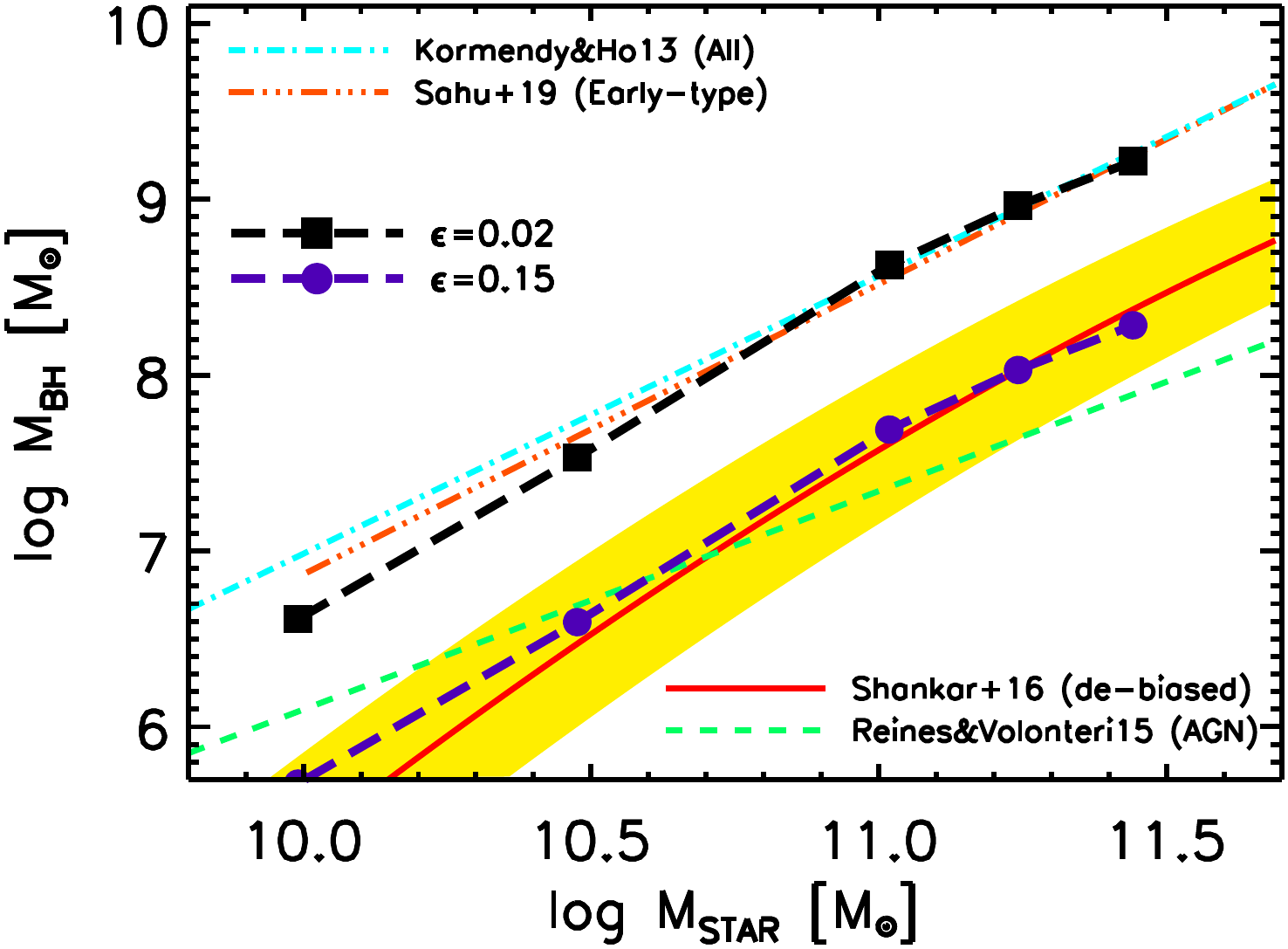}
    \caption{Correlations between central black hole mass and host galaxy \emph{total} stellar mass in the local Universe. The triple dot-dashed orange line is the fit to the local quiescent sample of early-type galaxies with dynamical measures of black holes by \citet{Sahu19}. The dot-dashed, cyan line is a linear fit to the sample of \citet{KormendyHo}. The solid red line with its scatter (yellow region) is the de-biased \mbh-\mstar\ relation from \citet{Shankar16BH}. The green dashed line is the fit to the local AGN from \citet{ReinesVolonteri15}. Also included are the predicted average black hole mass as a function of host stellar mass at $z=0.1$ for two different values of the radiative efficiency \epsi, as labelled. Values of $\epsie \sim 0.02$ are required (black long-dashed with filled squares) to match the normalization of the raw black hole \mbh-\mstar\ relation for local dynamically-measured quiescent black holes. A value of $\epsie \gtrsim 0.1$ is required (purple long-dashed with filled circles) to match the much lower \mbh-\mstar\ relation inferred from AGN or the de-biased relation of \citet{Shankar16BH}.
    \label{fig|FigureMbhMstar}}}
\end{figure*}

In what follows we will \emph{conservatively} adopt as a reference the stellar mass growth tracks derived by \citet{Moster18}, noticing that our core conclusions would be similar, in fact strengthened, by switching to any other semi-empirical model among those reported in the left panel of \figu\ref{fig|FigureMstarz}. The \citet{Grylls19b} model, in particular, predicts steeper
stellar mass growth histories which, at any given epoch, would correspond to moderately lower black hole accretion rates, which on average increase with host galaxy stellar mass \citep{Yang18}. Steeper stellar mass growth histories would thus naturally lead to lower cumulative black hole masses and a proportionally lower normalization in the accreted \mbh-\mstar\ relation, at fixed radiative/kinetic efficiencies, bolometric correction, and obscured fraction. In turn, to match the raw \mbh-\mstar\ local relation, these steeper models would require mean radiative efficiencies lower than those, already quite extreme (see \sect\ref{sec|results}), implied by the \citet{Moster18} stellar mass growth curves.

The right panel of \figu\ref{fig|FigureMstarz} shows the implied black hole mass accretion histories $\langle \mbhe \rangle(\mstare[z],z)$ obtained from direct time integration of the black hole accretion rates $\langle \macce \rangle(\mstare[z],z)$, included in \figu\ref{fig|FigureBHacc}, assuming a negligible kinetic efficiency and a nominal radiative efficiency of $\epsie=0.1$. As mentioned above, we adopt the stellar mass growth tracks by \citet{Moster18}, and assume an initial black hole mass at $z=4$ of $\mstare/10^4$, sufficiently small to have a minor impact on the mass accreted at later epochs. As discussed by \citet{Yang18}, the choice of initial black hole mass has an overall negligible effect on the cumulative black hole masses at $z\lesssim 1.5-2$. The growth histories exhibit ``downsizing'' -  a shift towards growth of lower mass black holes at later times - which broadly mirrors the one in stellar mass reported in the left panel of \figu\ref{fig|FigureMstarz}. We stress that the connection between black hole and stellar mass growth in \figu\ref{fig|FigureMstarz} does not necessarily imply any causal connection between the two.

\begin{figure*}
    \center{\includegraphics[width=\textwidth]{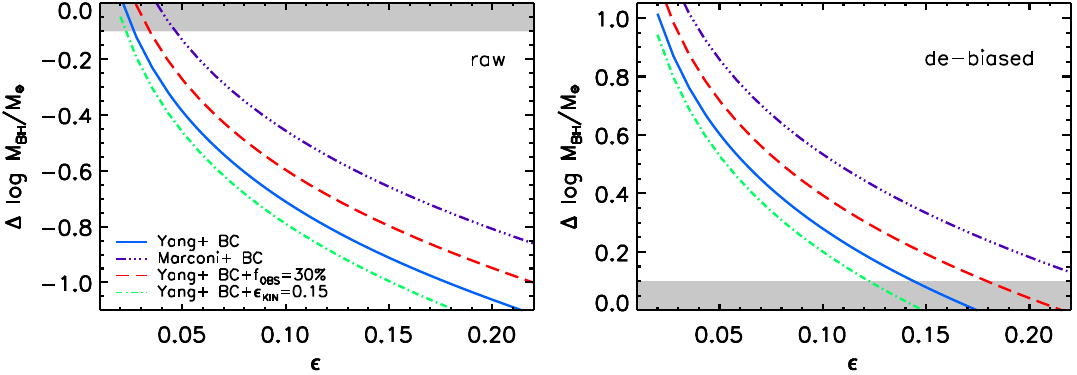}
    \caption{\emph{Left}: Displacement $\Delta \log \mbhe$ between the $\log \mbhe-\log \mstare$ relations of \citet{KormendyHo} and the one inferred from direct integration of the black hole accretion rate. \emph{Right}: Displacement in $\log \mbhe$ between the $\log \mbhe-\log \mstare$ relations of \citet{Shankar16BH} and the one inferred from direct integration of the black hole accretion rate. The solid blue, long-dashed red, triple dot-dashed purple, and green dot-dashed lines refer, respectively, to models based on the bolometric correction by \citet{Yang18}, on the bolometric correction by \citet{Marconi04}, the bolometric correction by \citet{Yang18} plus some correction for obscured sources, and the bolometric correction by \citet{Yang18} plus a kinetic efficiency of $\epsie_{\rm kin}=0.15$. Higher bolometric corrections or significant obscured fractions require larger radiative efficiencies to reproduce the de-biased \mbh-\mstar\ relation.
    \label{fig|FigureSummary}}}
\end{figure*}

\figu\ref{fig|FigureMBHz} depicts the ratio of the average black hole and stellar mass growth evolutionary histories, along the progenitor tracks $\mstare[z]$ (left panel), and at fixed stellar mass (right panel), as labelled. It is interesting to see that, first off, the ratio $\langle\mbhe[z]\rangle/\langle\mstare[z]\rangle$ is not constant for all galaxies but it steadily decreases with decreasing stellar mass by up to an order of magnitude. Second, all ratios irrespective of redshift or stellar mass lie below the average black hole-to-stellar mass ratio inferred locally by \citet[][cyan region]{KormendyHo}. Third, all $\langle\mbhe[z]\rangle/\langle\mstare[z]\rangle$ ratios tend to remain roughly constant up until at least $z\sim 2$ at fixed stellar mass, in line with a number of previous studies, obtained via Monte Carlo approaches \citep{Fiore17}, continuity equation models \citep{ZhangLu12,ShankarMsigma,Shankar13acc}, integration of the star formation rate (DelVecchio et al. 2019, submitted) or direct observations
\citep[e.g.,][see also \citealt{Suh19}]{GaskellMbhSigma,Salviander13,Shen15}, all suggesting weak evolution in the black hole-galaxy scaling relations. On the other hand, the $\langle\mbhe[z]\rangle/\langle\mstare[z]\rangle$ ratios may tend to decrease at high redshifts, though this trend may be sensitive to the exact choice of initial black hole masses chosen at $z\gtrsim 4$, especially relevant in lower mass systems.

\subsection{The comparison with the local \mbh-\mstar\ relation: Towards constraining the mean radiative efficiency $\epsie$}
\label{subsec|compareLocalMbhMstar}

Having devised robust methods to compute average stellar and black hole masses at any relevant epoch, we can compute the \mbh-\mstar\ relation in particular at $z\sim 0.1$ to compare with that independently inferred from local dynamical measures of supermassive black holes. \figu\ref{fig|FigureMbhMstar} reports the latest renditions of the \mbh-\mstar\ relation. All data sets in \figu\ref{fig|FigureMbhMstar} have been adjusted to the mass-to-light ratios adopted by \citet{Shankar16BH}, based on \citet{Bell03SEDs}. We first apply a linear fit to the \citet{KormendyHo} local inactive sample dynamically-measured supermassive black holes, as included in Table 3 of \citet{ReinesVolonteri15}, and correct stellar masses following \eq A1 in \citet{Shankar19BH}. The orange, triple dot-dashed line shows the linear fit by \citet{Sahu19} to early-type galaxies, where we conservatively set the parameter $v=1$ in their \eq11\ (lower values of $v$, as suggested by \citealt{Davis18}, would result in even higher normalizations). The raw \mbh-\mstar\ relation by \citet{Savorgnan16}, not reported in \figu\ref{fig|FigureMbhMstar},
is in broad agreement with the \citet{KormendyHo} relation \citep{Shankar19Nat}\footnote{For completeness, as already discussed by \citet{Shankar19Nat}, we also note that \citet{Davis18} have recently inferred a \mbh-\mstar\ relation for local dynamically measured black holes hosted in late-type galaxies significantly steeper than the one by \citet{Sahu19}, roughly consistent with the \citet{Shankar16BH} estimate at $\log \mstare/\msune\sim 10.5$, but rapidly approaching the \citet{Sahu19} relation at $\log \mstare/\msune\gtrsim 11$.}. The dashed green line shows the \mbh-\mstar\ relation inferred from single-epoch black hole mass estimates for AGN host galaxies by \citet[][see also \citealt{Baron19,Shankar19BH}]{ReinesVolonteri15}, assuming a mean virial parameters $\fvire=4.3$. In our terminology, the \citet{KormendyHo} and \citet{Sahu19} relations are ``raw'' estimates that fit the dynamically estimated black hole masses without accounting for the fact that this observed subset may be biased by the requirement of resolving the sphere of influence. The AGN sample is not subject to this bias, and the inferred \mbh-\mstar\ relation is about an order-of-magnitude below the raw relations for inactive black holes at $\mstare\sim 10^{11}\, \msune$.

As introduced in \sect\ref{sec|intro}, \citet[][see also \citealt{Shankar17BH,Shankar19BH}]{Shankar16BH} confirmed earlier claims \citep{Bernardi07} that black hole mass predominantly correlates with central stellar velocity dispersion \sis, with all other scaling relations with black hole mass being mostly driven by the former one. We focus here on the \mbh-\mstar\ relation because the higher redshift mean accretion rates are available as a function of stellar mass \citep{Yang18} rather than \sis, which is more difficult to measure. Using mock black hole samples that follow an \mbh-\sis\ relation and the \sis-\mstar\ relation of SDSS early-type galaxies, \citet{Shankar16BH} derived a de-biased \mbh-\mstar\ relation, valid for galaxies with $\mstare \gtrsim 2\times 10^{10}\, \msune$, which is shown by the red curve in \figu\ref{fig|FigureMbhMstar}, with the yellow band showing the inferred 1\sis\ scatter of \mbh\ at fixed \mstar. Including contributions from later-type galaxies would tend to produce slightly lower normalizations of the global unbiased \mbh-\mstar\ relation \citep[][]{Shankar19BH}. In principle the discrepancy between the raw \mbh-\mstar relations for quiescent black holes and the \citet{ReinesVolonteri15} result for AGN could arise because active galaxies have lower mass black holes, or because the virial factors used by \citet{ReinesVolonteri15} are much too low. However, a more natural interpretation of \figu\ref{fig|FigureMbhMstar} is that the de-biased \mbh-\mstar\ relation of \citet{Shankar16BH} is a better tracer of the mean \mbh-\mstar\ scaling relation, that active galaxies host black holes similar to those of other galaxies with the same stellar mass, and that virial factors are in line with theoretical expectations and empirically constrained models of the broad line region . This argument and its implications are explored in greater detail by \citet{Shankar19BH}.

\figu\ref{fig|FigureMbhMstar} presents an entirely independent argument for this point of view. Reproducing the raw \mbh-\mstar\ relation with our empirically based models of \sect\ref{subsec|aveMdotBH} requires a radiative efficiency $\epsie \sim 0.02$ (black dashed curve), well below the value expected from accretion disk theory \citep[e.g.,][]{SSdisk,Abramo13}. Reproducing the de-biased or AGN relation requires $\epsie \sim 0.15$ (purple dashed curve), in good agreement with theoretical predictions for accretion onto spinning black holes. This agreement between the de-biased local \mbh-\mstar\ relation and the prediction of a theoretically motivated, empirically based model is the principal result of this paper.


\subsection{The impact of systematics and robustness of results}
\label{subsec|systematics}

Although empirically based, our strategy still relies on a few input parameters and/or assumptions. In this \sect\ we will detail how our main results are robust against sensible variations of such inputs.
First off, the masses of supermassive black holes obtained by direct integration of \eq\ref{eq|averageMacc} require specification of the bolometric correction. Following \citet{Yang18}, in \figu\ref{fig|FigureMbhMstar} we have adopted as a reference the bolometric correction determined by \citet{Lusso12}. Other bolometric corrections proposed in the literature are characterized by up to factor of $\sim 3$ higher normalizations \citep[e.g.,][]{Marconi04,Hopkins06LF}. This will proportionally increase the integrated emissivity of AGN and thus the predicted final black hole mass at fixed stellar mass. Lining up to the same local \mbh-\mstar\ relation will therefore require a nearly proportional increase in the mean radiative efficiency \epsi, as one can see from the appearance of the ratio $\kbe/\epsie$ in \eq\ref{eq|averageMacc}. Another important point is that the average X-ray luminosities adopted in \eq\ref{eq|averageMacc} and taken from \citet{Yang18} do not necessarily account for possible additional large populations of hidden Compton-thick AGN. If present, the latter would clearly increase the total intrinsic X-ray luminosities and thus the black hole accretion rates and predicted final masses, at fixed stellar mass, bolometric correction, and radiative efficiency. On the other hand, allowing for a non-negligible kinetic efficiency $\epsie_{\rm kin}$ as expected from studies of radio-loud AGN \citep{MerloniHeinz07,Shankar08Cav,LaFranca10,Ghisellini13,Zubo18}, would tend to decrease the required mean radiative efficiency, when fixing the other parameters.

We summarise these behaviours in \figu\ref{fig|FigureSummary}. The solid blue, triple dot-dashed purple, long-dashed red, and green dot-dashed lines refer, respectively, to models based (see \eq\ref{eq|averageMacc}) on the bolometric correction by \citet{Yang18}, on the bolometric correction by \citet{Marconi04}, on the bolometric correction by \citet{Yang18} plus an additional multiplicative factor of 1.3 in \eq\ref{eq|averageMacc} to account for possible underestimates of the total mean intrinsic X-ray luminosity due to missed Compton-thich AGN \citep[e.g.,][]{Ueda14,Harrison16,GeoAkilas19,Ana19}, and on the bolometric correction \citet{Yang18} plus a kinetic efficiency of $\epsie_{\rm kin}=0.15$. The left panel shows the displacement, at a reference stellar mass of $\log \mstare/\msune=11$, in $\log \mbhe$ between the $\log \mbhe-\log \mstare$ relation of \citet{KormendyHo} and the one inferred from direct integration of the black hole accretion rate. The right panel shows the same quantity for the de-biased relation of \citet{Shankar16BH}. Adopting the higher bolometric correction or the additional 30\% obscured accretion fraction increases the implied radiative efficiency, but we would still require $\epsie\lesssim 0.04$ to reproduce the raw \mbh-\mstar\ relation to within 0.1 dex. The latest recalibration of the hard X-ray AGN bolometric corrections (Duras et al. submitted) tends to disfavour ``higher'' bolometric corrections \citep{Marconi04,Hop07} and well align with those determined by \citet{Lusso12}.

\begin{figure}
    \center{\includegraphics[width=\columnwidth]{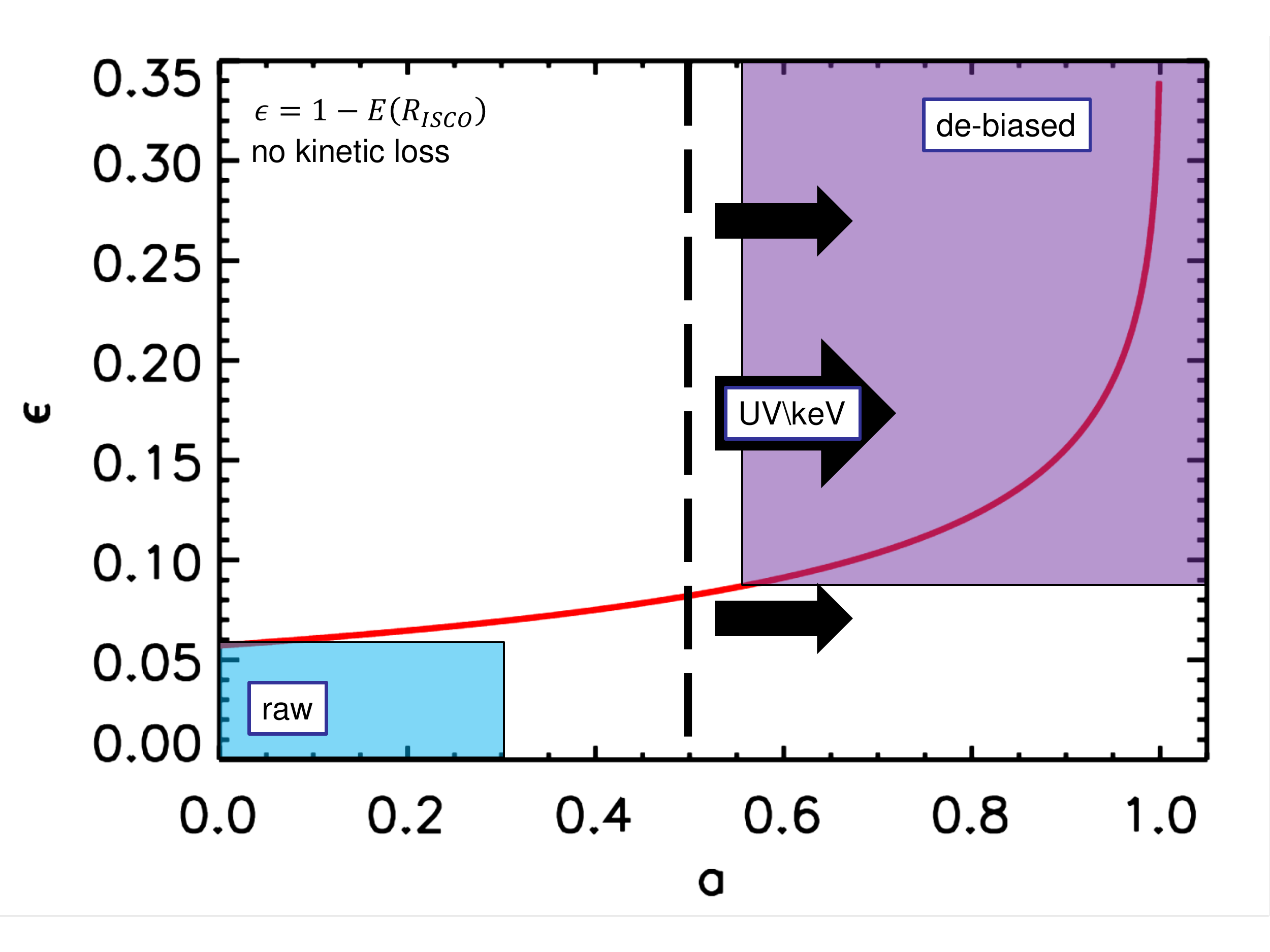}
    \caption{Radiative efficiency as a function of black hole spin (solid, red line) for direct accretion assuming no kinetic losses and $\varepsilon= 1-E$ with $E$ the energy at the
innermost stable circular orbit\citet{Bardeen72}. The constraints on the mean radiative efficiency arising from the fit to the intrinsic/unbiased \mbh-\mstar\ relation, $\varepsilon \gtrsim 0.1$ (with dimensionless spin parameter $a\gtrsim 0.5$), are shown with a purple rectangle, while those from the observed \mbh-\mstar\ relation, $\varepsilon \lesssim 0.05$, are shown with a cyan rectangle. The independent estimates of the spin parameter from UV/X-ray spectral modelling (black arrows) are broadly consistent with the former estimates with $a \gtrsim 0.5$.
    \label{fig|FigureEpsilonSummary}}}
\end{figure}

It is worth emphasizing that throughout this work we are, by design, dealing with mean radiative efficiencies modelled via the thin-disc approximation \citep{SSdisk}. Broad distributions of radiative efficiencies are indeed expected \citep[e.g.,][]{ZhangLu19eta}. In particular, substantial portions of the black hole population accreting at very low radiative efficiencies could be missed in our modelling. As discussed by \citet[][and references therein]{Yang18}, very low radiative efficiencies, significantly below the thin-disc approximation, for example in ADAF-like states, are expected to become effective only in extremely low Eddington ratio regimes (below 1\% of the Eddington limit). Such accretion mode is however too slow to provide a visible contribution to the final black holes, building mass on e-folding timescales much longer than the Hubble time \citep[see][for further details]{Yang18}.

\section{Conclusions}
\label{sec|discu}

As sketched in \figu\ref{fig|SummaryCartoon}, we have put forward a complementary approach to the classical \citet{Soltan} method, taking advantage of recent measurements of the average X-ray luminosity of accreting black holes as a function of galaxy stellar mass and redshift \citep{Yang18}, and of recent empirical models for the evolution of galaxy stellar masses \citep{Moster18}. For an assumed mean radiative efficiency \epsi, these empirical inputs allow us to predict the mean \mbh-\mstar\ relation as a function of redshift. We focus on the mass range $\log \mstare/\msune \sim 10.5-11.2$, where mergers are expected to be minor contributors to stellar mass and black hole growth \citep[e.g.,][and references therein]{Shankar13acc,Lapi18,Moster18}. Assuming constant radiative efficiency, we infer (\figu\ref{fig|FigureMBHz}) that the normalization and shape of the \mbh-\mstar\ relation is nearly independent of redshift at least up to $z\sim 2$, in agreement with the findings of \citet{Yang18}. Weak or negligible evolution of the \mbh-\sis\ relation has been inferred from analysis based on the black hole continuity equation \citep{ShankarMsigma} and from some direct observational studies \citep{GaskellMbhSigma,Shen15}. A non-evolving \mbh-\mstar\ relation implies that the stellar masses and central black hole masses of galaxies grow, on average, at the same rate over cosmic time. A non-evolving \mbh-\mstar\ and \mbh-\sis\ relations would also imply weak evolutions in the \sis-\mstar\ relations and in the overall fundamental plane of massive galaxies and their central black holes (see, e.g., discussion in \citealt{Suh19}).

Most importantly, we find (\figu\ref{fig|FigureMbhMstar}) that reproducing the raw observed relation between galaxy stellar masses and the dynamically inferred masses of inactive black holes requires a radiative efficiency $\epsie \sim 0.02$, well below theoretical expectations for thin accretion disks and values inferred from UV spectral energy distribution fitting \citep[e.g.,][]{DavisLaor11,Benny14,Cape15,Shankar16} and X-ray reflection analysis \citep{Reynolds13}. Higher bolometric corrections or significant fractions of obscured accretion can increase the inferred \epsi, but we still find $\epsie \lesssim 0.05$ for reasonable assumptions about these uncertainties (\figu\ref{fig|FigureSummary}). This mismatch between the inferred $\epsie$ and physical expectations provides independent evidence that the raw \mbh-\mstar\ relation for inactive black holes is biased high because black hole masses are only measured when the radius of gravitational influence is resolved, as argued by \citet{Shankar16BH}. Using \citet{Shankar16BH}'s de-biased \mbh-\mstar\ relation, or the relation inferred from AGN black hole mass estimates by \citet{ReinesVolonteri15}, we find a mean radiative efficiency $\epsie \sim 0.15$, in good agreement with theoretical expectations for accretion onto black holes with spin parameters $a \sim 0.5-1$. The red solid line in \figu\ref{fig|FigureEpsilonSummary} shows the monotonic dependence of the spin parameter on radiative efficiency, obtained by integrating the specific energy and orbital angular momentum equations in the limit\citep{Bardeen72,ZhangLu19} of no kinetic loss $\epsie=1-E(R_{\rm isco})$, with $E(R_{\rm isco})$ the specific orbital energy at the innermost stable circular orbit with radius $R_{\rm isco}$. This model suggests that values of $a \gtrsim 0.5$ would correspond to radiative efficiencies greater than $\epsie \gtrsim 0.1$ (black arrows in \figu\ref{fig|FigureEpsilonSummary}), which would be in line with the limits on \epsi\ obtained in this work when comparing with the de-biased \mbh-\mstar\ relations (purple area in the upper right of \figu\ref{fig|FigureEpsilonSummary}), but in tension with the allowed ranges of \epsi\ required by the match to the raw \mbh-\mstar\ relations (cyan area in the bottom left of \figu\ref{fig|FigureEpsilonSummary}). Flux limit effects may bias current X-ray surveys towards higher luminosity sources, possibly characterized by larger radiative efficiencies/spins \citep[][]{Vasudevan16}. As mentioned in \citet{Shankar19Nat}, the lower limits on the current AGN X-ray samples map to black holes radiating down to minimal radiative efficiencies of $\epsie\sim 0.05$ and accreting at $\gtrsim 10\%$ the Eddington limit, well within the thin-disc limit during which most of the final black hole mass is expected to assemble \citep{Yang18}.

Uncertainties in bolometric corrections, kinetic feedback efficiency, and other observational inputs are large enough that we cannot clearly rule out efficiencies $\epsilon < 0.1$ achievable with non-spinning black holes, though models would still require relatively high $\epsie_{\rm kin}$ to accommodate very low radiative efficiencies (\figu\ref{fig|FigureSummary}). A non-negligible obscured AGN fraction $f$ for galaxies in our stellar mass range \citep[e.g.,][]{Harrison16,Ana19} would increase our inferred $\epsie$ by a factor $\sim (1+f)$, so at face value our results favour $\epsie \gtrsim 0.15-0.20$, implying high characteristic spin parameters $a\gtrsim 0.9$. Most direct measurements of black hole spins from X-ray reflection spectroscopy favour $a\gtrsim 0.5$ (see, for example, Table 1 in \citealt{ZhangLu19}), a finding further corroborated by UV spectral energy distribution modelling \citep[e.g.,][]{Cape15,Shankar16}. Future observations and modelling can reduce uncertainties in bolometric corrections and the contribution of obscured accretion. They can also test our predictions against direct observations of the (non)-evolving \mbh-\mstar\
relation (Suh et al. 2019, submitted), AGN and quasar clustering \citep[e.g.,][]{Shankar10shen}, and the cross-correlation of AGN and galaxies \citep[e.g.,][]{Krumpe15}.



\section*{Acknowledgments}

FS acknowledges Peter Behroozi for sharing his stellar mass accretion tracks. FS acknowledges partial support from a Leverhulme Trust Research Fellowship. RC acknowledges financial support from CONICYT Doctorado Nacional N$^\circ$\,21161487 and CONICYT PIA ACT172033.
DMA thanks the Science and Technology Facilities Council (STFC) for support from grant ST/L00075X/1. MM acknowledges support from the Beatriu de Pinos fellowship (2017-BP-00114).

\appendix
\section{The impact of redshift and aperture}
\label{Appendix}

\begin{figure}
    \center{\includegraphics[width=\columnwidth]{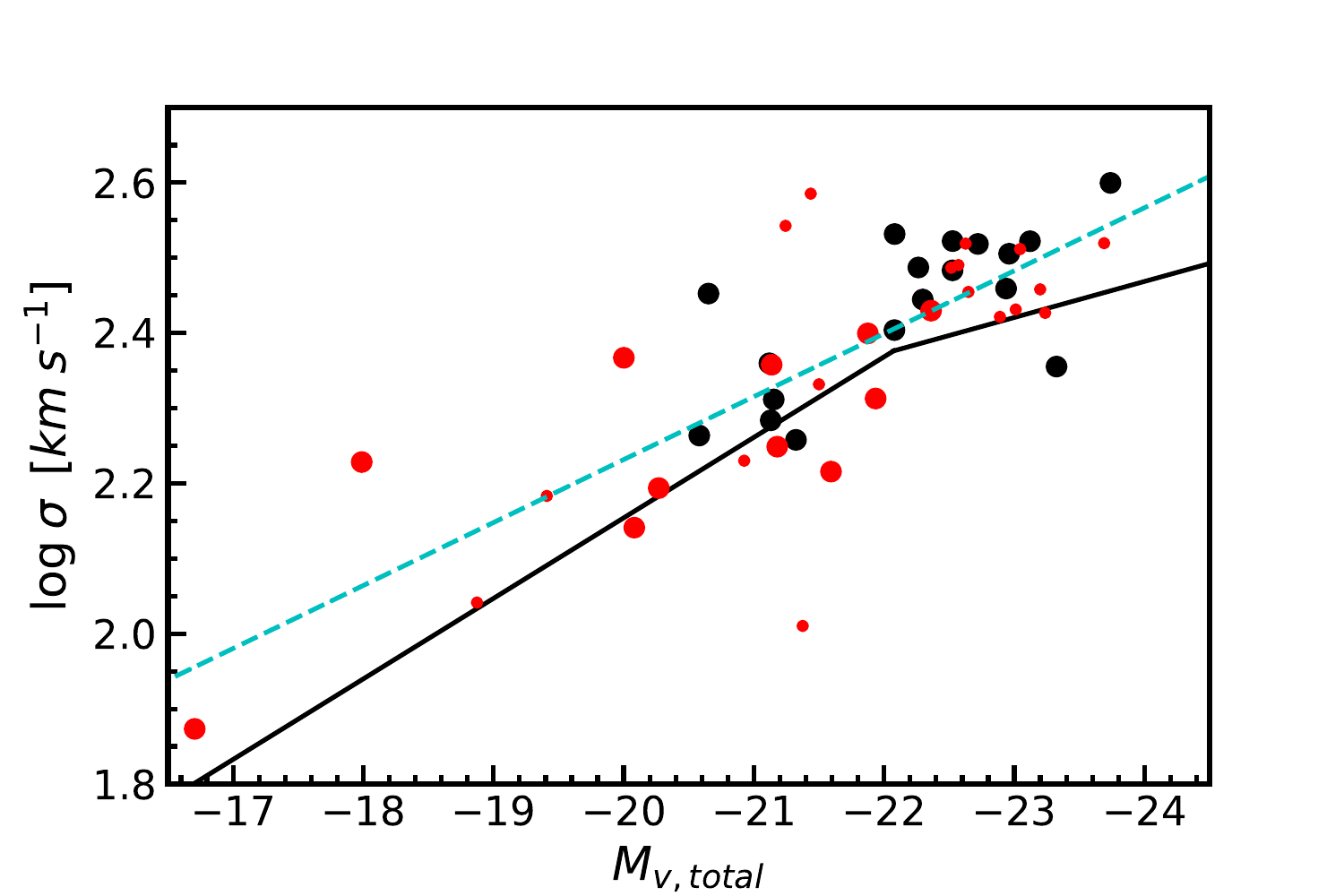}
    \caption{Same layout as \figu2 in \citet{Kormendy19}.
    Comparison between the double power-law local \sis-$M_V$ relation (solid black line) by \citet{Lauer07demo} and \citet{KormendyBender13}, with the \citet{Kormendy19} data set of galaxies with dynamical measurements of their central supermassive black hole mass subdivided into core (black circles) and coreless (red circles) galaxies. The long-dashed, cyan line is a linear fit to the
    \citet{Kormendy19} black hole data, proving that at fixed (total) galaxy magnitude $M_V$, local black holes' hosts tend to have larger mean velocity dispersions than the underlying population of local galaxies.
    \label{fig|FigureKormendy}}}
\end{figure}

In a very recent proceeding of the IAU Symposium 2019, \citet{Kormendy19} stated that local scaling relations of dynamically measured black holes are not biased.
In this Appendix we carefully address this statement in light of his data and addressing some of his concerns.
In our assessment, the bias in the black hole scaling relations that we have identified in our previous papers is fully consistent with the data recently presented by \citet{Kormendy19}.

\citet{Kormendy19} first of all notices that the host galaxies of dynamically measured black holes follow the same scaling relations traced by larger serendipitous samples of local galaxies (e.g., his \figu1). \citet{Shankar16BH,Shankar17BH} have indeed demonstrated that, when compared to local SDSS galaxies, most scaling relations in terms of effective radius, S\'{e}rsic index and dynamical mass are very similar for galaxies with and without central black hole dynamical mass measurement. \citet{Shankar16BH,Shankar17BH,Shankar19BH},
following \citet{Bernardi07}, highlighted that the bias is mostly evident in the velocity dispersion distributions at fixed stellar mass. The hosts of supermassive black holes tend to have mean velocity dispersions, on average, systematically higher by $\sim 0.05-0.2$ dex, with the discrepancy gradually increasing towards lower stellar masses, than SDSS galaxies. Although this discrepancy is apparently relatively small, it could generate offsets in mean black hole masses up to a factor of $\sim 2-10$, on the assumptions that black hole mass is primarily related to velocity dispersion scaling as $\mbhe\propto \sise^5$, as suggested by residuals analysis \citep{Bernardi07,Shankar16BH,Shankar17BH,Shankar19BH} and the study of mono- and bivariate correlations \citep{deNicolaMarconi19}. It is important to note that the analysis of \citet{Shankar16BH,Shankar17BH} is based on the \citet{Savorgnan16} sample of \emph{early-type} galaxies with dynamical black hole mass measurement, which was for full consistency compared with only early-type SDSS galaxies, with minimal contribution from pseudobulges \citet{Shankar16BH,Shankar17BH}. \citet{Shankar19BH} further extended the comparison between spirals in the SDSS galaxies and the (few) spirals in the \citet{Savorgnan16} sample, showing that for black holes hosted in spirals the bias in mean velocity dispersion at fixed total stellar mass persists but it is less evident.

\citet{Kormendy19} attempts in his \figu2 a similar comparison between velocity dispersion and total galaxy magnitude in $V$ band for local galaxies with black holes, and the broken power-law velocity dispersion-absolute magnitude scaling relation of local galaxies from \citet{Lauer07demo} and \citet{KormendyBender13}. We
propose a similar comparison in \figu\ref{fig|FigureKormendy} in which we linearly fit his sample of galaxies with black holes (cyan long-dashed line) and compare it with his quoted velocity dispersion-absolute magnitude relation (black solid line). It is apparent that the mean velocity dispersion in the \citet{Kormendy19} black hole sample still presents an offset of $\sim 0.05-0.2$ dex at fixed galaxy magnitude, increasing with decreasing galaxy luminosity. Indeed \citet{Kormendy19} recognizes that velocity dispersions in his sample tend to lie above the mean velocity dispersion-absolute magnitude relation of local galaxies, and also addresses the issue of incompleteness in the local sample of black holes, as more distant galaxies have not been searched for.
The offset between the two relations in \figu\ref{fig|FigureKormendy} appears small, but it is enough to cause a large bias in the \mbh-\mstar\ relation because the \mbh-\sis\ relation is so steep.

Last but not least, \citet{Kormendy19} highlights the possible bias inherent in the SDSS survey dominated by more distant galaxies. Fixed apertures would naturally sample larger radii of the galaxies and possibly measure lower velocity dispersions at fixed stellar mass. To check for this possible aperture-distance effect we have analysed ~2000 early-type galaxies in MAnGA with IFU spectroscopy. We have seen that indeed velocity dispersions appear slightly larger at very low redshifts $z\lesssim 0.04$ than at $z\gtrsim 0.2$ for galaxies with $\mstare \sim 10^{11}\msune$, but it is negligible for galaxies $\mstare \lesssim 3\sim 10^{10}\msune$, in which instead the bias in velocity dispersion discussed above should be evident.

We also note that \citet{Kormendy19} does not mention the increasing sample of serendipitous local AGN, inclusive of early and late-type galaxies \citep[e.g.,][]{Busch15,ReinesVolonteri15,Baron19},
that tend to lie up to an order of magnitude below the \mbh-\mstar\ relation of inactive, dynamically measured local black holes, providing further, independent evidence of the bias in the latter sample \citep{Shankar19BH}.

We conclude that the bias in black hole scaling relations that we infer in this paper (and our previous papers) is consistent with the data presented by \citet{Kormendy19}.

\bibliographystyle{mn2e_Daly}
\bibliography{MNRASMbhMsNEWViola}

\begin{thebibliography}{116}
\expandafter\ifx\csname natexlab\endcsname\relax\def\natexlab#1{#1}\fi

\bibitem[{{Abramowicz} \& {Fragile}(2013)}]{Abramo13}
{Abramowicz} M.~A., {Fragile} P.~C., 2013, Living Reviews in Relativity, 16, 1

\bibitem[{{Ananna} {et~al}\mbox{.}(2019){Ananna}, {Treister}, {Urry}, {Ricci},
  {Kirkpatrick}, {LaMassa}, {Buchner}, {Civano}, {Tremmel}, \&
  {Marchesi}}]{Ana19}
{Ananna} T.~T. {et~al.}, 2019, \apj, 871, 240

\bibitem[{{Aversa} {et~al}\mbox{.}(2015){Aversa}, {Lapi}, {de Zotti},
  {Shankar}, \& {Danese}}]{Aversa15}
{Aversa} R., {Lapi} A., {de Zotti} G., {Shankar} F., {Danese} L., 2015, \apj,
  810, 74

\bibitem[{{Barausse} {et~al}\mbox{.}(2017){Barausse}, {Shankar}, {Bernardi},
  {Dubois}, \& {Sheth}}]{Barausse17}
{Barausse} E., {Shankar} F., {Bernardi} M., {Dubois} Y., {Sheth} R.~K., 2017,
  \mnras, 468, 4782

\bibitem[{{Bardeen} {et~al}\mbox{.}(1972){Bardeen}, {Press}, \&
  {Teukolsky}}]{Bardeen72}
{Bardeen} J.~M., {Press} W.~H., {Teukolsky} S.~A., 1972, \apj, 178, 347

\bibitem[{{Baron} \& {M{\'e}nard}(2019)}]{Baron19}
{Baron} D., {M{\'e}nard} B., 2019, arXiv e-prints, arXiv:1903.01996

\bibitem[{{Batcheldor} {et~al}\mbox{.}(2007){Batcheldor}, {Marconi}, {Merritt},
  \& {Axon}}]{Batcheldor07}
{Batcheldor} D., {Marconi} A., {Merritt} D., {Axon} D.~J., 2007, \apjl, 663,
  L85

\bibitem[{{Behroozi} {et~al}\mbox{.}(2018){Behroozi}, {Wechsler}, {Hearin}, \&
  {Conroy}}]{Behroozi18}
{Behroozi} P., {Wechsler} R., {Hearin} A., {Conroy} C., 2018, ArXiv e-prints

\bibitem[{{Bell} {et~al}\mbox{.}(2003){Bell}, {McIntosh}, {Katz}, \&
  {Weinberg}}]{Bell03SEDs}
{Bell} E.~F., {McIntosh} D.~H., {Katz} N., {Weinberg} M.~D., 2003, \apjs, 149,
  289

\bibitem[{{Bentz} \& {Manne-Nicholas}(2018)}]{Bentz18}
{Bentz} M.~C., {Manne-Nicholas} E., 2018, ArXiv e-prints

\bibitem[{{Bentz} {et~al}\mbox{.}(2008){Bentz}, {Peterson}, {Netzer}, {Pogge},
  \& {Vestergaard}}]{Bentz08RL}
{Bentz} M.~C., {Peterson} B.~M., {Netzer} H., {Pogge} R.~W., {Vestergaard} M.,
  2008, ArXiv e-prints

\bibitem[{{Bernardi} {et~al}\mbox{.}(2013){Bernardi}, {Meert}, {Sheth},
  {Vikram}, {Huertas-Company}, {Mei}, \& {Shankar}}]{Bernardi13}
{Bernardi} M., {Meert} A., {Sheth} R.~K., {Vikram} V., {Huertas-Company} M.,
  {Mei} S., {Shankar} F., 2013, \mnras, 436, 697

\bibitem[{{Bernardi} {et~al}\mbox{.}(2007){Bernardi}, {Sheth}, {Tundo}, \&
  {Hyde}}]{Bernardi07}
{Bernardi} M., {Sheth} R.~K., {Tundo} E., {Hyde} J.~B., 2007, \apj, 660, 267

\bibitem[{{Bruzual} \& {Charlot}(2003)}]{BruzualCharlot03}
{Bruzual} G., {Charlot} S., 2003, \mnras, 344, 1000

\bibitem[{{Busch} {et~al}\mbox{.}(2016){Busch}, {Fazeli}, {Eckart},
  {Valencia-S.}, {Smaji{\'c}}, {Moser}, {Scharw{\"a}chter}, {Dierkes}, \&
  {Fischer}}]{Busch15}
{Busch} G. {et~al.}, 2016, \aap, 587, A138

\bibitem[{{Busch} {et~al}\mbox{.}(2014){Busch}, {Zuther}, {Valencia-S.},
  {Moser}, {Fischer}, {Eckart}, {Scharw{\"a}chter}, {Gadotti}, \&
  {Wisotzki}}]{busch2014low}
{Busch} G. {et~al.}, 2014, \aap, 561, A140

\bibitem[{{Capellupo} {et~al}\mbox{.}(2015{\natexlab{a}}){Capellupo}, {Netzer},
  {Lira}, {Trakhtenbrot}, \& {Mej{\'{\i}}a-Restrepo}}]{Cape15}
{Capellupo} D.~M., {Netzer} H., {Lira} P., {Trakhtenbrot} B.,
  {Mej{\'{\i}}a-Restrepo} J., 2015{\natexlab{a}}, \mnras, 446, 3427

\bibitem[{{Capellupo} {et~al}\mbox{.}(2015{\natexlab{b}}){Capellupo}, {Netzer},
  {Lira}, {Trakhtenbrot}, \& {Mej{\'{\i}}a-Restrepo}}]{Capellupo15}
{Capellupo} D.~M., {Netzer} H., {Lira} P., {Trakhtenbrot} B.,
  {Mej{\'{\i}}a-Restrepo} J., 2015{\natexlab{b}}, \mnras, 446, 3427

\bibitem[{{Chabrier}(2003)}]{Chabrier03}
{Chabrier} G., 2003, \pasp, 115, 763

\bibitem[{{Dasyra} {et~al}\mbox{.}(2007){Dasyra}, {Tacconi}, {Davies},
  {Genzel}, {Lutz}, {Peterson}, {Veilleux}, {Baker}, {Schweitzer}, \&
  {Sturm}}]{Dasyra07}
{Dasyra} K.~M. {et~al.}, 2007, \apj, 657, 102

\bibitem[{{Davidzon} {et~al}\mbox{.}(2017){Davidzon}, {Ilbert}, {Laigle},
  {Coupon}, {McCracken}, {Delvecchio}, {Masters}, {Capak}, {Hsieh}, {Le
  F{\`e}vre}, {Tresse}, {Bethermin}, {Chang}, {Faisst}, {Le Floc'h},
  {Steinhardt}, {Toft}, {Aussel}, {Dubois}, {Hasinger}, {Salvato}, {Sanders},
  {Scoville}, \& {Silverman}}]{Davidzon17}
{Davidzon} I. {et~al.}, 2017, \aap, 605, A70

\bibitem[{{Davis} {et~al}\mbox{.}(2018){Davis}, {Graham}, \&
  {Cameron}}]{Davis18}
{Davis} B.~L., {Graham} A.~W., {Cameron} E., 2018, \apj, 869, 113

\bibitem[{{Davis} \& {Laor}(2011)}]{DavisLaor11}
{Davis} S.~W., {Laor} A., 2011, \apj, 728, 98

\bibitem[{{de Nicola} {et~al}\mbox{.}(2019){de Nicola}, {Marconi}, \&
  {Longo}}]{deNicolaMarconi19}
{de Nicola} S., {Marconi} A., {Longo} G., 2019, arXiv e-prints,
  arXiv:1909.01749

\bibitem[{{Fabian}(2012)}]{Fabian12}
{Fabian} A.~C., 2012, \araa, 50, 455

\bibitem[{{Falomo} {et~al}\mbox{.}(2014){Falomo}, {Bettoni}, {Karhunen},
  {Kotilainen}, \& {Uslenghi}}]{Falomo14}
{Falomo} R., {Bettoni} D., {Karhunen} K., {Kotilainen} J.~K., {Uslenghi} M.,
  2014, \mnras, 440, 476

\bibitem[{{Ferrarese} \& {Ford}(2005)}]{FerrareseFord}
{Ferrarese} L., {Ford} H., 2005, Space Science Reviews, 116, 523

\bibitem[{{Ferrarese} \& {Merritt}(2000)}]{Ferrarese00}
{Ferrarese} L., {Merritt} D., 2000, \apjl, 539, L9

\bibitem[{{Fiore} {et~al}\mbox{.}(2017){Fiore}, {Feruglio}, {Shankar},
  {Bischetti}, {Bongiorno}, {Brusa}, {Carniani}, {Cicone}, {Duras}, {Lamastra},
  {Mainieri}, {Marconi}, {Menci}, {Maiolino}, {Piconcelli}, {Vietri}, \&
  {Zappacosta}}]{Fiore17}
{Fiore} F. {et~al.}, 2017, \aap, 601, A143

\bibitem[{{Gaskell}(2009)}]{GaskellMbhSigma}
{Gaskell} C.~M., 2009, ArXiv:0908.0328

\bibitem[{{Georgantopoulos} \& {Akylas}(2019)}]{GeoAkilas19}
{Georgantopoulos} I., {Akylas} A., 2019, \aap, 621, A28

\bibitem[{{Ghisellini} {et~al}\mbox{.}(2013){Ghisellini}, {Haardt}, {Della
  Ceca}, {Volonteri}, \& {Sbarrato}}]{Ghisellini13}
{Ghisellini} G., {Haardt} F., {Della Ceca} R., {Volonteri} M., {Sbarrato} T.,
  2013, \mnras, 432, 2818

\bibitem[{{Graham}(2016)}]{GrahamReview15}
{Graham} A.~W., 2016, Galactic Bulges, 418, 263

\bibitem[{{Graham} \& {Scott}(2015)}]{GrahamScott15}
{Graham} A.~W., {Scott} N., 2015, \apj, 798, 54

\bibitem[{{Granato} {et~al}\mbox{.}(2004){Granato}, {De Zotti}, {Silva},
  {Bressan}, \& {Danese}}]{Granato04}
{Granato} G.~L., {De Zotti} G., {Silva} L., {Bressan} A., {Danese} L., 2004,
  \apj, 600, 580

\bibitem[{{Greene} {et~al}\mbox{.}(2016){Greene}, {Seth}, {Kim}, {L{\"a}sker},
  {Goulding}, {Gao}, {Braatz}, {Henkel}, {Condon}, {Lo}, \&
  {Zhao}}]{greene2016megamaser}
{Greene} J.~E. {et~al.}, 2016, \apjl, 826, L32

\bibitem[{{Grier} {et~al}\mbox{.}(2017){Grier}, {Pancoast}, {Barth},
  {Fausnaugh}, {Brewer}, {Treu}, \& {Peterson}}]{Grier17}
{Grier} C.~J., {Pancoast} A., {Barth} A.~J., {Fausnaugh} M.~M., {Brewer} B.~J.,
  {Treu} T., {Peterson} B.~M., 2017, \apj, 849, 146

\bibitem[{{Grylls} {et~al}\mbox{.}(2019{\natexlab{a}}){Grylls}, {Shankar},
  {Leja}, {Menci}, {Moster}, {Behroozi}, \& {Zanisi}}]{Grylls19b}
{Grylls} P.~J., {Shankar} F., {Leja} J., {Menci} N., {Moster} B., {Behroozi}
  P., {Zanisi} L., 2019{\natexlab{a}}, \mnras, 2560

\bibitem[{{Grylls} {et~al}\mbox{.}(2019{\natexlab{b}}){Grylls}, {Shankar},
  {Zanisi}, \& {Bernardi}}]{Grylls19}
{Grylls} P.~J., {Shankar} F., {Zanisi} L., {Bernardi} M., 2019{\natexlab{b}},
  \mnras, 483, 2506

\bibitem[{{G{\"u}ltekin} {et~al}\mbox{.}(2011){G{\"u}ltekin}, {Tremaine},
  {Loeb}, \& {Richstone}}]{Gultekin11}
{G{\"u}ltekin} K., {Tremaine} S., {Loeb} A., {Richstone} D.~O., 2011, \apj,
  738, 17

\bibitem[{{H{\"a}ring} \& {Rix}(2004)}]{HaringRix}
{H{\"a}ring} N., {Rix} H.-W., 2004, \apjl, 604, L89

\bibitem[{{Harrison} {et~al}\mbox{.}(2016){Harrison}, {Aird}, {Civano},
  {Lansbury}, {Mullaney}, {Ballantyne}, {Alexander}, {Stern}, {Ajello},
  {Barret}, {Bauer}, {Balokovi{\'c}}, {Brandt}, {Brightman}, {Boggs},
  {Christensen}, {Comastri}, {Craig}, {Del Moro}, {Forster}, {Gandhi},
  {Giommi}, {Grefenstette}, {Hailey}, {Hickox}, {Hornstrup}, {Kitaguchi},
  {Koglin}, {Luo}, {Madsen}, {Mao}, {Miyasaka}, {Mori}, {Perri}, {Pivovaroff},
  {Puccetti}, {Rana}, {Treister}, {Walton}, {Westergaard}, {Wik}, {Zappacosta},
  {Zhang}, \& {Zoglauer}}]{Harrison16}
{Harrison} F.~A. {et~al.}, 2016, \apj, 831, 185

\bibitem[{{Hopkins} {et~al}\mbox{.}(2008){Hopkins}, {Hernquist}, {Cox}, \&
  {Kere{\v s}}}]{Hop08clust}
{Hopkins} P.~F., {Hernquist} L., {Cox} T.~J., {Kere{\v s}} D., 2008, \apjs,
  175, 356

\bibitem[{{Hopkins} {et~al}\mbox{.}(2006){Hopkins}, {Hernquist}, {Cox},
  {Robertson}, {Di Matteo}, \& {Springel}}]{Hopkins06LF}
{Hopkins} P.~F., {Hernquist} L., {Cox} T.~J., {Robertson} B., {Di Matteo} T.,
  {Springel} V., 2006, \apj, 639, 700

\bibitem[{{Hopkins} {et~al}\mbox{.}(2007){Hopkins}, {Richards}, \&
  {Hernquist}}]{Hop07}
{Hopkins} P.~F., {Richards} G.~T., {Hernquist} L., 2007, \apj, 654, 731

\bibitem[{{Kim} {et~al}\mbox{.}(2008){Kim}, {Ho}, {Peng}, {Barth}, {Im},
  {Martini}, \& {Nelson}}]{Kim08}
{Kim} M., {Ho} L.~C., {Peng} C.~Y., {Barth} A.~J., {Im} M., {Martini} P.,
  {Nelson} C.~H., 2008, \apj, 687, 767

\bibitem[{{King}(2003)}]{King03}
{King} A., 2003, \apjl, 596, L27

\bibitem[{{Kormendy}(2019)}]{Kormendy19}
{Kormendy} J., 2019, arXiv e-prints, arXiv:1909.10821

\bibitem[{{Kormendy} \& {Bender}(2013)}]{KormendyBender13}
{Kormendy} J., {Bender} R., 2013, \apjl, 769, L5

\bibitem[{{Kormendy} \& {Ho}(2013)}]{KormendyHo}
{Kormendy} J., {Ho} L.~C., 2013, \araa, 51, 511

\bibitem[{{Krumpe} {et~al}\mbox{.}(2015){Krumpe}, {Miyaji}, {Husemann},
  {Fanidakis}, {Coil}, \& {Aceves}}]{Krumpe15}
{Krumpe} M., {Miyaji} T., {Husemann} B., {Fanidakis} N., {Coil} A.~L., {Aceves}
  H., 2015, \apj, 815, 21

\bibitem[{{La Franca} {et~al}\mbox{.}(2010){La Franca}, {Melini}, \&
  {Fiore}}]{LaFranca10}
{La Franca} F., {Melini} G., {Fiore} F., 2010, \apj, 718, 368

\bibitem[{{Laigle} {et~al}\mbox{.}(2016){Laigle}, {McCracken}, {Ilbert},
  {Hsieh}, {Davidzon}, {Capak}, {Hasinger}, {Silverman}, {Pichon}, {Coupon},
  {Aussel}, {Le Borgne}, {Caputi}, {Cassata}, {Chang}, {Civano}, {Dunlop},
  {Fynbo}, {Kartaltepe}, {Koekemoer}, {Le F{\`e}vre}, {Le Floc'h}, {Leauthaud},
  {Lilly}, {Lin}, {Marchesi}, {Milvang-Jensen}, {Salvato}, {Sanders},
  {Scoville}, {Smolcic}, {Stockmann}, {Taniguchi}, {Tasca}, {Toft}, {Vaccari},
  \& {Zabl}}]{Laigle16}
{Laigle} C. {et~al.}, 2016, \apjs, 224, 24

\bibitem[{{Lapi} {et~al}\mbox{.}(2018){Lapi}, {Pantoni}, {Zanisi}, {Shi},
  {Mancuso}, {Massardi}, {Shankar}, {Bressan}, \& {Danese}}]{Lapi18}
{Lapi} A. {et~al.}, 2018, \apj, 857, 22

\bibitem[{{Lapi} {et~al}\mbox{.}(2006){Lapi}, {Shankar}, {Mao}, {Granato},
  {Silva}, {De Zotti}, \& {Danese}}]{Lapi06}
{Lapi} A., {Shankar} F., {Mao} J., {Granato} G.~L., {Silva} L., {De Zotti} G.,
  {Danese} L., 2006, \apj, 650, 42

\bibitem[{{L{\"a}sker} {et~al}\mbox{.}(2014){L{\"a}sker}, {Ferrarese}, {van de
  Ven}, \& {Shankar}}]{Laesker14}
{L{\"a}sker} R., {Ferrarese} L., {van de Ven} G., {Shankar} F., 2014, \apj,
  780, 70

\bibitem[{{Lauer} {et~al}\mbox{.}(2007){Lauer}, {Faber}, {Richstone},
  {Gebhardt}, {Tremaine}, {Postman}, {Dressler}, {Aller}, {Filippenko},
  {Green}, {Ho}, {Kormendy}, {Magorrian}, \& {Pinkney}}]{Lauer07demo}
{Lauer} T.~R. {et~al.}, 2007, \apj, 662, 808

\bibitem[{{Lusso} {et~al}\mbox{.}(2012){Lusso}, {Comastri}, {Simmons},
  {Mignoli}, {Zamorani}, {Vignali}, {Brusa}, {Shankar}, {Lutz}, {Trump},
  {Maiolino}, {Gilli}, {Bolzonella}, {Puccetti}, {Salvato}, {Impey}, {Civano},
  {Elvis}, {Mainieri}, {Silverman}, {Koekemoer}, {Bongiorno}, {Merloni},
  {Berta}, {Le Floc'h}, {Magnelli}, {Pozzi}, \& {Riguccini}}]{Lusso12}
{Lusso} E. {et~al.}, 2012, \mnras, 425, 623

\bibitem[{{Lynden-Bell}(1969)}]{Lynden69}
{Lynden-Bell} D., 1969, \nat, 223, 690

\bibitem[{{Magorrian} {et~al}\mbox{.}(1998){Magorrian}, {Tremaine},
  {Richstone}, {Bender}, {Bower}, {Dressler}, {Faber}, {Gebhardt}, {Green},
  {Grillmair}, {Kormendy}, \& {Lauer}}]{Magorrian98}
{Magorrian} J. {et~al.}, 1998, \aj, 115, 2285

\bibitem[{{Maraston}(2005)}]{Maraston05}
{Maraston} C., 2005, Monthly Notices of the Royal Astronomical Society, 362,
  799

\bibitem[{{Marconi} {et~al}\mbox{.}(2004){Marconi}, {Risaliti}, {Gilli},
  {Hunt}, {Maiolino}, \& {Salvati}}]{Marconi04}
{Marconi} A., {Risaliti} G., {Gilli} R., {Hunt} L.~K., {Maiolino} R., {Salvati}
  M., 2004, \mnras, 351, 169

\bibitem[{{Meert} {et~al}\mbox{.}(2015){Meert}, {Vikram}, \&
  {Bernardi}}]{Meert15}
{Meert} A., {Vikram} V., {Bernardi} M., 2015, \mnras, 446, 3943

\bibitem[{{Merloni} \& {Heinz}(2007)}]{MerloniHeinz07}
{Merloni} A., {Heinz} S., 2007, \mnras, 381, 589

\bibitem[{{Morabito} \& {Dai}(2012)}]{DaiMbhSigma}
{Morabito} L.~K., {Dai} X., 2012, \apj, 757, 172

\bibitem[{{Moster} {et~al}\mbox{.}(2018){Moster}, {Naab}, \&
  {White}}]{Moster18}
{Moster} B.~P., {Naab} T., {White} S.~D.~M., 2018, \mnras, 477, 1822

\bibitem[{{Moster} {et~al}\mbox{.}(2019){Moster}, {Naab}, \&
  {White}}]{Moster19}
{Moster} B.~P., {Naab} T., {White} S. D.~M., 2019, arXiv e-prints,
  arXiv:1910.09552

\bibitem[{{Mullaney} {et~al}\mbox{.}(2012){Mullaney}, {Daddi}, {B{\'e}thermin},
  {Elbaz}, {Juneau}, {Pannella}, {Sargent}, {Alexander}, \& {Hickox}}]{Mulla12}
{Mullaney} J.~R. {et~al.}, 2012, \apjl, 753, L30

\bibitem[{{Ni} {et~al}\mbox{.}(2019){Ni}, {Yang}, {Brandt}, {Alexander},
  {Chen}, {Luo}, {Vito}, \& {Xue}}]{Ni19}
{Ni} Q., {Yang} G., {Brandt} W.~N., {Alexander} D.~M., {Chen} C. T.~J., {Luo}
  B., {Vito} F., {Xue} Y.~Q., 2019, \mnras, 2248

\bibitem[{{Pancoast} {et~al}\mbox{.}(2014){Pancoast}, {Brewer}, \&
  {Treu}}]{Pancoast14}
{Pancoast} A., {Brewer} B.~J., {Treu} T., 2014, \mnras, 445, 3055

\bibitem[{{Peterson} {et~al}\mbox{.}(2004){Peterson}, {Ferrarese}, {Gilbert},
  {Kaspi}, {Malkan}, {Maoz}, {Merritt}, {Netzer}, {Onken}, {Pogge},
  {Vestergaard}, \& {Wandel}}]{Peterson04}
{Peterson} B.~M. {et~al.}, 2004, \apj, 613, 682

\bibitem[{{Rees}(1984)}]{Rees84}
{Rees} M.~J., 1984, \araa, 22, 471

\bibitem[{{Reines} \& {Volonteri}(2015)}]{ReinesVolonteri15}
{Reines} A.~E., {Volonteri} M., 2015, \apj, 813, 82

\bibitem[{{Reynolds}(2014)}]{Reynolds13}
{Reynolds} C.~S., 2014, \ssr, 183, 277

\bibitem[{{Ricci} {et~al}\mbox{.}(2017){Ricci}, {La Franca}, {Marconi},
  {Onori}, {Shankar}, {Schneider}, {Sani}, {Bianchi}, {Bongiorno}, {Brusa},
  {Fiore}, {Maiolino}, \& {Vignali}}]{Ricci17SR}
{Ricci} F. {et~al.}, 2017, \mnras, 471, L41

\bibitem[{{Rodriguez-Gomez} {et~al}\mbox{.}(2017){Rodriguez-Gomez}, {Sales},
  {Genel}, {Pillepich}, {Zjupa}, {Nelson}, {Griffen}, {Torrey}, {Snyder},
  {Vogelsberger}, {Springel}, {Ma}, \& {Hernquist}}]{Rodriguez17}
{Rodriguez-Gomez} V. {et~al.}, 2017, \mnras, 467, 3083

\bibitem[{{Sahu} {et~al}\mbox{.}(2019){Sahu}, {Graham}, \& {Davis}}]{Sahu19}
{Sahu} N., {Graham} A.~W., {Davis} B.~L., 2019, arXiv e-prints

\bibitem[{{Salucci} {et~al}\mbox{.}(1999){Salucci}, {Szuszkiewicz}, {Monaco},
  \& {Danese}}]{Salucci99}
{Salucci} P., {Szuszkiewicz} E., {Monaco} P., {Danese} L., 1999, \mnras, 307,
  637

\bibitem[{{Salviander} \& {Shields}(2013)}]{Salviander13}
{Salviander} S., {Shields} G.~A., 2013, \apj, 764, 80

\bibitem[{{Santini} {et~al}\mbox{.}(2015){Santini}, {Ferguson}, {Fontana},
  {Mobasher}, {Barro}, {Castellano}, {Finkelstein}, {Grazian}, {Hsu}, {Lee},
  {Lee}, {Pforr}, {Salvato}, {Wiklind}, {Wuyts}, {Almaini}, {Cooper},
  {Galametz}, {Weiner}, {Amorin}, {Boutsia}, {Conselice}, {Dahlen},
  {Dickinson}, {Giavalisco}, {Grogin}, {Guo}, {Hathi}, {Kocevski}, {Koekemoer},
  {Kurczynski}, {Merlin}, {Mortlock}, {Newman}, {Paris}, {Pentericci},
  {Simons}, \& {Willner}}]{Santini15}
{Santini} P. {et~al.}, 2015, The Astrophysical Journal, 801, 97

\bibitem[{{Sarria} {et~al}\mbox{.}(2010){Sarria}, {Maiolino}, {La Franca},
  {Pozzi}, {Fiore}, {Marconi}, {Vignali}, \& {Comastri}}]{Sarria10}
{Sarria} J.~E., {Maiolino} R., {La Franca} F., {Pozzi} F., {Fiore} F.,
  {Marconi} A., {Vignali} C., {Comastri} A., 2010, \aap, 522, L3

\bibitem[{{Savorgnan} {et~al}\mbox{.}(2016){Savorgnan}, {Graham}, {Marconi}, \&
  {Sani}}]{Savorgnan16}
{Savorgnan} G.~A.~D., {Graham} A.~W., {Marconi} A., {Sani} E., 2016, \apj, 817,
  21

\bibitem[{{Shakura} \& {Sunyaev}(1973)}]{SSdisk}
{Shakura} N.~I., {Sunyaev} R.~A., 1973, \aap, 24, 337

\bibitem[{{Shankar}(2009)}]{ShankarReview}
{Shankar} F., 2009, New Astron. Rev., 53, 57

\bibitem[{{Shankar} {et~al}\mbox{.}(2019{\natexlab{a}}){Shankar}, {Allevato},
  {Bernardi}, {Marsden}, {Lapi}, {Menci}, {Grylls}, {Krumpe}, {Zanisi},
  {Ricci}, {La Franca}, {Baldi}, {Moreno}, \& {Sheth}}]{Shankar19Nat}
{Shankar} F. {et~al.}, 2019{\natexlab{a}}, arXiv e-prints, arXiv:1910.10175

\bibitem[{{Shankar} {et~al}\mbox{.}(2009{\natexlab{a}}){Shankar}, {Bernardi},
  \& {Haiman}}]{ShankarMsigma}
{Shankar} F., {Bernardi} M., {Haiman} Z., 2009{\natexlab{a}}, \apj, 694, 867

\bibitem[{{Shankar} {et~al}\mbox{.}(2019{\natexlab{b}}){Shankar}, {Bernardi},
  {Richardson}, {Marsden}, {Sheth}, {Allevato}, {Graziani}, {Mezcua}, {Ricci},
  {Penny}, {La Franca}, \& {Pacucci}}]{Shankar19BH}
{Shankar} F. {et~al.}, 2019{\natexlab{b}}, \mnras

\bibitem[{{Shankar} {et~al}\mbox{.}(2017){Shankar}, {Bernardi}, \&
  {Sheth}}]{Shankar17BH}
{Shankar} F., {Bernardi} M., {Sheth} R.~K., 2017, \mnras

\bibitem[{{Shankar} {et~al}\mbox{.}(2016{\natexlab{a}}){Shankar}, {Bernardi},
  {Sheth}, {Ferrarese}, {Graham}, {Savorgnan}, {Allevato}, {Marconi},
  {L{\"a}sker}, \& {Lapi}}]{Shankar16BH}
{Shankar} F. {et~al.}, 2016{\natexlab{a}}, \mnras, 460, 3119

\bibitem[{{Shankar} {et~al}\mbox{.}(2016{\natexlab{b}}){Shankar}, {Calderone},
  {Knigge}, {Matthews}, {Buckland}, {Hryniewicz}, {Sivakoff}, {Dai},
  {Richardson}, {Riley}, {Gray}, {La Franca}, {Altamirano}, {Croston},
  {Gandhi}, {H{\"o}nig}, {McHardy}, \& {Middleton}}]{Shankar16}
{Shankar} F. {et~al.}, 2016{\natexlab{b}}, \apjl, 818, L1

\bibitem[{{Shankar} {et~al}\mbox{.}(2008){Shankar}, {Cavaliere}, {Cirasuolo},
  \& {Maraschi}}]{Shankar08Cav}
{Shankar} F., {Cavaliere} A., {Cirasuolo} M., {Maraschi} L., 2008, \apj, 676,
  131

\bibitem[{{Shankar} {et~al}\mbox{.}(2006){Shankar}, {Lapi}, {Salucci}, {De
  Zotti}, \& {Danese}}]{Shankar06}
{Shankar} F., {Lapi} A., {Salucci} P., {De Zotti} G., {Danese} L., 2006, \apj,
  643, 14

\bibitem[{{Shankar} {et~al}\mbox{.}(2013{\natexlab{a}}){Shankar}, {Marulli},
  {Bernardi}, {Mei}, {Meert}, \& {Vikram}}]{Shankar13}
{Shankar} F., {Marulli} F., {Bernardi} M., {Mei} S., {Meert} A., {Vikram} V.,
  2013{\natexlab{a}}, \mnras, 428, 109

\bibitem[{{Shankar} {et~al}\mbox{.}(2004){Shankar}, {Salucci}, {Granato}, {De
  Zotti}, \& {Danese}}]{Shankar04}
{Shankar} F., {Salucci} P., {Granato} G.~L., {De Zotti} G., {Danese} L., 2004,
  \mnras, 354, 1020

\bibitem[{{Shankar} {et~al}\mbox{.}(2009{\natexlab{b}}){Shankar}, {Weinberg},
  \& {Miralda-Escud{\'e}}}]{SWM}
{Shankar} F., {Weinberg} D.~H., {Miralda-Escud{\'e}} J., 2009{\natexlab{b}},
  \apj, 690, 20

\bibitem[{{Shankar} {et~al}\mbox{.}(2013{\natexlab{b}}){Shankar}, {Weinberg},
  \& {Miralda-Escud{\'e}}}]{Shankar13acc}
{Shankar} F., {Weinberg} D.~H., {Miralda-Escud{\'e}} J., 2013{\natexlab{b}},
  \mnras, 428, 421

\bibitem[{{Shankar} {et~al}\mbox{.}(2010){Shankar}, {Weinberg}, \&
  {Shen}}]{Shankar10shen}
{Shankar} F., {Weinberg} D.~H., {Shen} Y., 2010, \mnras, 406, 1959

\bibitem[{{Shen} {et~al}\mbox{.}(2015){Shen}, {Greene}, {Ho}, {Brandt},
  {Denney}, {Horne}, {Jiang}, {Kochanek}, {McGreer}, {Merloni}, {Peterson},
  {Petitjean}, {Schneider}, {Schulze}, {Strauss}, {Tao}, {Trump}, {Pan}, \&
  {Bizyaev}}]{Shen15}
{Shen} Y. {et~al.}, 2015, \apj, 805, 96

\bibitem[{{Silk} \& {Rees}(1998)}]{SilkRees}
{Silk} J., {Rees} M.~J., 1998, \aap, 331, L1

\bibitem[{{Soltan}(1982)}]{Soltan}
{Soltan} A., 1982, \mnras, 200, 115

\bibitem[{{Suh} {et~al}\mbox{.}(2019){Suh}, {Civano}, {Trakhtenbrot},
  {Shankar}, {Hasinger}, {Sanders}, \& {Allevato}}]{Suh19}
{Suh} H., {Civano} F., {Trakhtenbrot} B., {Shankar} F., {Hasinger} G.,
  {Sanders} D.~B., {Allevato} V., 2019, arXiv:1912.02824, arXiv:1912.02824

\bibitem[{{Thorne}(1974)}]{Thorne74}
{Thorne} K.~S., 1974, \apj, 191, 507

\bibitem[{{Trakhtenbrot}(2014)}]{Benny14}
{Trakhtenbrot} B., 2014, \apjl, 789, L9

\bibitem[{{Ueda} {et~al}\mbox{.}(2014){Ueda}, {Akiyama}, {Hasinger}, {Miyaji},
  \& {Watson}}]{Ueda14}
{Ueda} Y., {Akiyama} M., {Hasinger} G., {Miyaji} T., {Watson} M.~G., 2014,
  \apj, 786, 104

\bibitem[{{van den Bosch} {et~al}\mbox{.}(2015){van den Bosch}, {Gebhardt},
  {G{\"u}ltekin}, {Y{\i}ld{\i}r{\i}m}, \& {Walsh}}]{Remco15}
{van den Bosch} R.~C.~E., {Gebhardt} K., {G{\"u}ltekin} K., {Y{\i}ld{\i}r{\i}m}
  A., {Walsh} J.~L., 2015, \apjs, 218, 10

\bibitem[{{Vasudevan} {et~al}\mbox{.}(2016){Vasudevan}, {Fabian}, {Reynolds},
  {Aird}, {Dauser}, \& {Gallo}}]{Vasudevan16}
{Vasudevan} R.~V., {Fabian} A.~C., {Reynolds} C.~S., {Aird} J., {Dauser} T.,
  {Gallo} L.~C., 2016, \mnras, 458, 2012

\bibitem[{{Yang} {et~al}\mbox{.}(2019){Yang}, {Brandt}, {Alexander}, {Chen},
  {Ni}, {Vito}, \& {Zhu}}]{Yang19}
{Yang} G., {Brandt} W.~N., {Alexander} D.~M., {Chen} C. T.~J., {Ni} Q., {Vito}
  F., {Zhu} F.~F., 2019, \mnras, 485, 3721

\bibitem[{{Yang} {et~al}\mbox{.}(2018){Yang}, {Brandt}, {Vito}, {Chen},
  {Trump}, {Luo}, {Sun}, {Xue}, {Koekemoer}, {Schneider}, {Vignali}, \&
  {Wang}}]{Yang18}
{Yang} G. {et~al.}, 2018, \mnras, 475, 1887

\bibitem[{{Yang} {et~al}\mbox{.}(2017){Yang}, {Chen}, {Vito}, {Brandt},
  {Alexander}, {Luo}, {Sun}, {Xue}, {Bauer}, {Koekemoer}, {Lehmer}, {Liu},
  {Schneider}, {Shemmer}, {Trump}, {Vignali}, \& {Wang}}]{Yang17}
{Yang} G. {et~al.}, 2017, \apj, 842, 72

\bibitem[{{Yu} \& {Lu}(2008)}]{YuLu08}
{Yu} Q., {Lu} Y., 2008, \apj, 689, 732

\bibitem[{{Yu} \& {Tremaine}(2002)}]{YuTremaine}
{Yu} Q., {Tremaine} S., 2002, \mnras, 335, 965

\bibitem[{{Zhang} \& {Lu}(2017)}]{ZhangLu19eta}
{Zhang} X., {Lu} Y., 2017, Science China Physics, Mechanics, and Astronomy, 60,
  109511

\bibitem[{{Zhang} \& {Lu}(2019)}]{ZhangLu19}
{Zhang} X., {Lu} Y., 2019, \apj, 873, 101

\bibitem[{{Zhang} {et~al}\mbox{.}(2012){Zhang}, {Lu}, \& {Yu}}]{ZhangLu12}
{Zhang} X., {Lu} Y., {Yu} Q., 2012, \apj, 761, 5

\bibitem[{{Zubovas}(2018)}]{Zubo18}
{Zubovas} K., 2018, \mnras, 479, 3189

\bibitem[{{Zubovas} \& {King}(2019)}]{Zubovas19}
{Zubovas} K., {King} A.~R., 2019, General Relativity and Gravitation, 51, 65

\end{thebibliography}

\label{lastpage}
\end{document}